\documentclass[a4paper,fleqn,usenatbib]{mnras}
\usepackage{mathptmx}
\usepackage[T1]{fontenc}
\usepackage{ae,aecompl}
\usepackage{deluxetable}
\usepackage{pdflscape}
\usepackage{multicol,multirow}
\usepackage{verbatim}
\usepackage{color}

\usepackage{graphicx}	
\usepackage{amsmath}	
\usepackage{amssymb}	
\usepackage{multirow,multicol}

\newcommand{\xmm}{\hbox{\it XMM-Newton\/}}
\newcommand{\chandra}{{\it Chandra\/}}
\newcommand{\flux}{{erg~cm$^{-2}$~s$^{-1}$}}
\newcommand{\mflux}{{erg~cm$^{-2}$~s$^{-1}$~Hz$^{-1}$}}
\newcommand{\lum}{{erg~s$^{-1}$}}
\newcommand{\mlum}{{erg~s$^{-1}$~Hz$^{-1}$}}
\newcommand\iona[2]{#1$\;${\scshape{#2}}}
\def\xray{{\hbox{X-ray}}}
\def\xrays{{\hbox{X-rays}}}
\def\aox{$\alpha_{\rm OX}$}
\def\daox{$\Delta\alpha_{\rm OX}$}
\defcitealias{Luo2015}{Luo15}
\defcitealias{Wu2011}{Wu11}
\defcitealias{Wu2012}{Wu12}

\title[Connecting weak-line and typical quasars]{Connecting the X-ray properties of weak-line 
and typical quasars: testing for a geometrically thick accretion disk}

\author[Q. Ni et al.]{Q. Ni,$^{1}$\thanks{E-mail: qxn1@psu.edu}
W.~N.~Brandt,$^{1,2,3}$
B.~Luo,$^{4,5}$
P.~B.~Hall,$^{6}$
Yue~Shen,$^{7}$
S.~F.~Anderson,$^{8}$
\newauthor
R.~M.~Plotkin,$^{9}$
Gordon~T.~Richards,$^{10}$
D.~P.~Schneider,$^{1,2}$
O.~Shemmer$^{11}$
and J.~Wu$^{12}$
\\
$^{1}$Department of Astronomy \& Astrophysics, Davey Lab, The Pennsylvania State University, University Park, PA 16802, USA\\
$^{2}$Institute for Gravitation and the Cosmos, The Pennsylvania State University, University Park, PA 16802, USA\\
$^{3}$Department of Physics, 104 Davey Lab, The Pennsylvania State University, University Park, PA 16802, USA\\
$^{4}$School of Astronomy and Space Science, Nanjing University, Nanjing, Jiangsu 210093, China\\
$^{5}$Key Laboratory of Modern Astronomy and Astrophysics (Nanjing University), Ministry of Education, Nanjing, Jiangsu 210093, China\\
$^{6}$Department of Physics \& Astronomy, York University, 4700 Keele Street, Toronto, ON M3J 1P3, Canada\\
$^{7}$Department of Astronomy, University of Illinois at Urbana-Champaign, Urbana, IL 61801, USA\\
$^{8}$Department of Astronomy, University of Washington, Box 351580, Seattle, WA 98195, USA\\
$^{9}$International Centre for Radio Astronomy Research -- Curtin University, GPO Box U1987, Perth, WA 6845, Australia\\
$^{10}$Department of Physics, Drexel University, 32 S. 32nd Street, Philadelphia, PA 19104, USA\\
$^{11}$Department of Physics, University of North Texas, Denton, TX 76203, USA\\
$^{12}$Department of Astronomy, Xiamen University, Xiamen, Fujian 361005, China
}

\date{Accepted XXX. Received YYY; in original form ZZZ}

\pubyear{2018}

\begin{document}
\label{firstpage}
\pagerange{\pageref{firstpage}--\pageref{lastpage}}
\maketitle

\begin{abstract}
We present \xray\ and multiwavelength analyses of 32 weak 
emission-line quasars (WLQs) selected in a consistent and unbiased 
manner. New \chandra\ \hbox{3.1--4.8~ks} observations were 
obtained for 14 of these WLQs with \iona{C}{iv} rest-frame equivalent widths (REWs) of 5--15~\AA, 
and these serve as an \xray\ observational ``bridge'' between 
previously studied WLQs with \iona{C}{iv} REW $\lesssim~$5~\AA\ 
and more-typical quasars with \iona{C}{iv} REW $\approx$ 15--100~\AA.
We have identified and quantified a strong dependence of the 
fraction of \xray\ weak quasars upon \iona{C}{iv} REW; this 
fraction declines by a factor of $\approx 13$ (from 
$\approx 44$\% to $\approx 3$\%) for \iona{C}{iv} REW 
ranging from \hbox{4--50~\AA}, and the rate of decline appears
particularly strong in the 10--20~\AA\ range. The dependence broadly
supports the proposed ``shielding'' model for WLQs, in which a 
geometrically and optically thick inner accretion disk, expected 
for a quasar accreting at a high Eddington ratio, both prevents 
ionizing EUV/\xray\ photons from reaching the high-ionization 
broad emission-line region and also sometimes blocks
the line-of-sight to the central \xray\ emitting region. 
This model is also supported by the hard average spectral shape 
of \xray\ weak WLQs (with a power-law effective photon index 
of $\Gamma_{\rm eff}=1.19^{+0.56}_{-0.45}$). Additionally, we have 
examined UV continuum/emission-line properties that might 
trace \xray\ weakness among WLQs, confirming that red 
UV continuum color is the most-effective tracer. 
\end{abstract}




\section{Introduction}
\subsection{The importance, X-ray properties, and current interpretation of weak-line quasars} \label{ssec_1_1}
Strong and broad line emission is a defining characteristic 
of quasar spectra in the optical and ultraviolet (UV). However, 
the Sloan Digital Sky Survey (SDSS) \citep{York2000} has discovered a remarkable 
population of $\approx 200$ type~1 quasars with strikingly weak or 
no emission lines \citep[e.g.][]{Fan1999,DS2009,Plotkin2010}. 
They have Ly$\alpha$+N\,{\sc v} rest-frame equivalent widths (REWs) $\lesssim 15$~\AA\ 
and/or C\,{\sc iv} REWs $\lesssim10$~\AA; these numbers represent $\gtrsim 3\sigma$
negative deviations from the REW distribution means with no 
corresponding positive ``tail'' to the distribution. Their C\,{\sc iv} 
and other high-ionization lines also often show remarkably large
blueshifts of \hbox{$\approx 2000-10000$~km~s$^{-1}$} (e.g. \citealt[]{Wu2011,Wu2012,Luo2015}, hereafter \citetalias{Wu2011}; \citetalias{Wu2012}; \citetalias{Luo2015}). The majority of these weak-line quasars (WLQs) are radio quiet and are {\it not\/} 
BL~Lacs (e.g. \citetalias{Wu2011}; \citetalias{Wu2012}; \citetalias{Luo2015}). Multiwavelength and multi-epoch studies also argue strongly 
that the weakness of the lines is not caused by line obscuration, 
gravitational lensing or microlensing effects, broad absorption 
line (BAL) effects, or radiatively inefficient accretion flows
(e.g. \citealt[]{DS2009,Shemmer2010,Lane2011}; \citetalias{Wu2011}; \citetalias{Wu2012}; \citetalias{Luo2015}; \citealt[]{Plotkin2015}).
The fraction of WLQs among quasars also appears to rise with 
redshift \citep[e.g.][]{DS2009,Luo2015,Banados2016}, perhaps due to generally higher Eddington 
ratios ($L/L_{\rm Edd}$) at high redshift (e.g. \citealt{SK2012}).

Investigations of the basic \xray\ properties 
of WLQs have found them to be remarkable objects
(e.g. \citetalias{Wu2011}; \citetalias{Wu2012}; \citetalias{Luo2015}). 
First, about half of the WLQ population shows notably 
weak \xray\ emission (by an average factor of $\approx 20$)
compared to expectations from their optical/UV continuum 
emission (i.e., considering \aox\ 
and \daox).\footnote{\label{aoxfootnote}\aox\ is the power-law slope 
connecting the monochromatic luminosities at rest-frame 2500~\AA\ and 2~keV; i.e., 
$\alpha_{\rm OX}=0.3838 \log(L_{\rm 2~keV}/L_{2500~\mathring{\rm{A}}})$.
This parameter is found to be correlated with $L_{2500~\mathring{\rm{A}}}$
\citep[e.g.][]{Wilkes1994,Vignali2003,Strateva2005,Steffen2006,Just2007,Lusso2010}. We also define 
$\Delta\alpha_{\rm OX}=\alpha_{\rm OX}({\rm Observed})-\alpha_{\rm OX}(L_{2500~\mathring{\rm{A}}})$, 
which quantifies the deviation of the observed \xray\ 
luminosity relative to that expected from the 
\aox-$L_{2500~\mathring{\rm{A}}}$ relation. \daox\ is used to derive the stated 
factors of \xray\ weakness.} This \xray\ weakness is further
confirmed by examination of the broad-band spectral energy
distributions (SEDs) of WLQs (e.g. see Section 4.2 of \citetalias{Luo2015}). The fraction of \xray\ weak 
objects among WLQs appears to be much higher than that for 
the general quasar population ($\lesssim 8$\% for \daox\ $<-0.2$, corresponding to an \xray\ weakness factor $>3.3$; 
e.g. \citealt{Gibson2008}), although quantification
of this effect has been challenging owing to the strong
selection effects in previous WLQ samples with sensitive
\xray\ coverage. 

Stacking of the counts from the \xray\ weak WLQs reveals that 
they have hard \xray\ spectra, on average, with effective power-law photon indices
of $\langle \Gamma \rangle\approx 1.4$ (\citetalias{Luo2015}). These hard \xray\ spectra suggest that high 
levels of intrinsic \xray\ absorption (with $N_{\rm H}$ of at least $10^{23}$~cm$^{-2}$, 
and perhaps much greater), Compton reflection, and/or scattering
 are commonly present, which is somewhat 
surprising given these quasars' typical blue UV/optical continua 
and broad, although weak, emission lines. The 
presence of such \xray\ absorption, reflection, and/or scattering is further 
supported by a \chandra\ spectrum of 
SDSS~J1521+5202, an extremely luminous WLQ showing 
$N_{\rm H}\gtrsim 1.3\times 10^{23}$~cm$^{-2}$ and allowing for Compton-reflection dominated spectral solutions
where the column density is $N_{\rm H}\gg 10^{24}$~cm$^{-2}$ (\citetalias{Luo2015}).
Finally, the WLQs in the half of the population that is not \xray\ 
weak show notably steep power-law spectra with 
$\langle \Gamma \rangle=2.2\pm 0.1$. Such steep power-law 
spectra generally indicate accretion at a high 
$L/L_{\rm Edd}$ 
\citep[e.g.][]{Shemmer2008,Risaliti2009,Brightman2013}. 

\citetalias{Luo2015} detailed a ``shielding'' model for 
WLQs that has the potential to explain, in a simple and unified 
manner, their weak UV lines, their remarkable \xray\ properties, 
and their other multiwavelength properties (also see \citetalias{Wu2011}, \citetalias{Wu2012}).
To explain the weak UV lines, this model relies upon small-scale 
($\lesssim 30R_{\rm S}$, where $R_{\rm S}$ is the Schwarzschild radius of the central black hole) 
shielding of ionizing EUV/\xray\ photons that prevents them from 
reaching the broad emission-line region (BELR). The shielding 
material is also responsible for the \xray\ weakness/absorption 
seen in about half of WLQs; as discussed below, the absorption 
may be highly Compton thick with 
$N_{\rm H}\gg 10^{24}$~cm$^{-2}$. 
When our line of sight 
intercepts the shield, we see an \xray\ weak WLQ; when it 
misses the shield, we observe an \xray\ normal WLQ. In both cases, ionizing EUV/\xray\ 
photons produced on small scales are mostly prevented from 
reaching the (largely equatorial and unobscured) high-ionization 
BELR, thereby causing the weak high-ionization line emission. 
However, UV/optical photons produced on larger scales in the 
accretion disk remain always unobscured. 

\begin{figure}
\begin{center}
\includegraphics[scale=0.355]{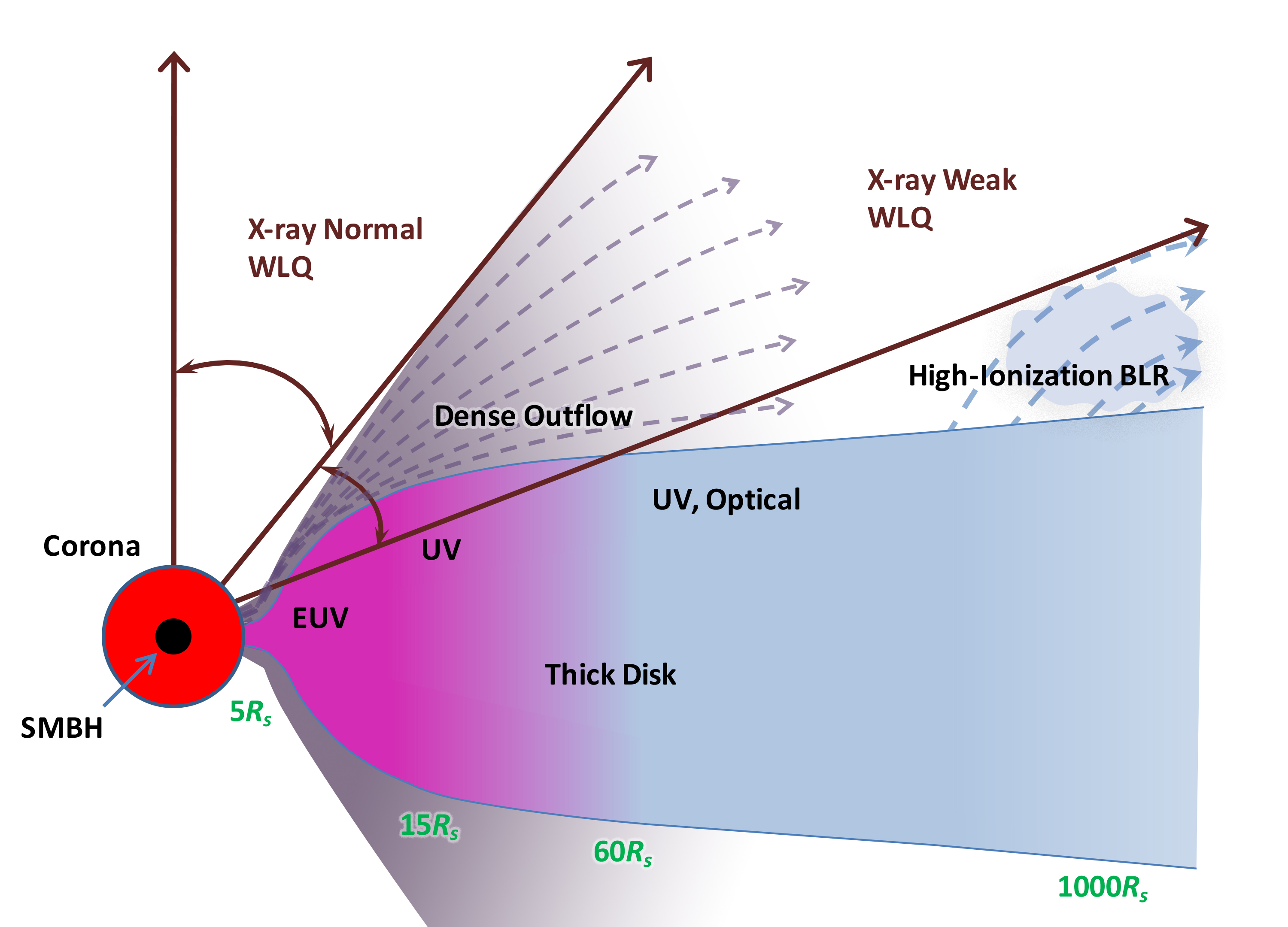}
\caption{
Schematic diagram of the geometrically thick-disk scenario for WLQs. 
The height of the disk increases with the Eddington ratio.
The geometrically thick disk (as well as the associated dense outflow) expected for a quasar accreting at high Eddington ratio prevents the nuclear ionizing photons
from reaching the high-ionization BELR, leading to the observed weak high-ionization line emission.
The observed \xray\ weakness occurs when our line of sight happens to intercept the thick disk.
The approximate radial scale of the disk is annotated in units of the Schwarzschild radius ($R_s$); the vertical extent of the disk beyond the outward flare of the profile at 5-15 $R_s$ is only shown schematically.}
\label{schematic}
\end{center}
\end{figure}

In the shielding model for WLQs, one must explain the nature 
of the putative shield---it must lie inside the BELR and be 
capable of blocking even high-energy \xrays. \citetalias{Luo2015}
proposed that the shield may be the geometrically and optically 
thick inner accretion disk expected for a quasar accreting at 
a high Eddington ratio 
\citep[e.g.][]{Abramowicz1988,Jiang2014,Jiang2017,Sadowski2014,Wang2014}. 
In Figure~\ref{schematic},
we present a schematic illustration of this shielding model based on Figure~18 of \citetalias{Luo2015}.\footnote{Our schematic in Figure~\ref{schematic}
differs from that in \citetalias{Luo2015} in the disk profile, the inclusion of the outflow, and the location of the BELR.
These changes have been made to reflect current thinking about super-Eddington accretion disks (e.g. S. Davis \& Y. Jiang 2015, personal communication; \citealt{Sadowski2015,Jiang2017}), in which optically thick outflows are expected and the radius at which the disk approaches a geometrically thin standard disk may be substantially larger than assumed in \citetalias{Luo2015}.}
A high Eddington 
ratio is consistent with the steep power-law \xray\ spectra 
seen for \xray\ normal WLQs as well as (admittedly uncertain)
$L/L_{\rm Edd}$ estimates based on virial black-hole masses
(\citetalias{Luo2015}; \citealt{Plotkin2015}). High $L/L_{\rm Edd}$ 
values are also consistent with arguments that the Eddington 
ratios of quasars generally increase with declining C\,{\sc iv} 
REWs and rising C\,{\sc iv} blueshifts 
\citep[e.g.][]{Shen2014,Sulentic2014,SL2015,Sun2018}.
We note that the thick disks associated with high $L/L_{\rm Edd}$ accretion
are expected to have substantial associated outflows
(e.g. \citealt{Jiang2014,Jiang2017,Dai2018}), and these likely
also contribute to the shielding (e.g. \citetalias{Wu2011,Wu2012}). However, the
detailed nature of these outflows remains uncertain. For this
reason, as well as the purpose of simplicity, throughout we
will just refer to the ``thick disk'' as the shield, and we
implicitly include any associated outflow within this term.

The continuity of the quasar population in the C\,{\sc iv} REW
vs.\ C\,{\sc iv} blueshift plane, as well as semi-analytic models and numerical simulations 
\citep[e.g.][]{Sadowski2009,Sadowski2011,Jiang2014,Jiang2017}, suggest that if a thick inner 
accretion disk is present in these extreme WLQs with high Eddington 
ratios, then it is also likely present, albeit with a lower height (and thus covering 
factor), in quasars with lower $L/L_{\rm Edd}$. Thus, shielding by 
inner disks may play a broader role, beyond just extreme WLQs, in 
regulating the high-ionization emission-line strengths of quasars
(e.g. \citealt{Leighly2004}; \citetalias{Luo2015}). As $L/L_{\rm Edd}$ continuously
declines within a sample of quasars, shielding may
also drop due to a reduction in the thickness of the inner accretion disk
allowing the production of generally stronger emission 
lines. Of course, other factors are also expected to influence 
emission-line strengths, including gas metallicity and anisotropic 
line emission. Due to object-to-object differences in these 
secondary factors, correlations between $L/L_{\rm Edd}$ and 
emission-line strengths are expected to have intrinsic scatter \citep[e.g.][]{SL2015}. 

While the above discussion has focused on the shielding model for WLQs, 
other explanations for their properties have also 
been put forward. These include 
an anemic BELR \citep[e.g.][]{Shemmer2010}, 
a still-developing BELR \citep[e.g.][]{Hryniewicz2010}, and 
a ``soft'' ionizing continuum arising from the relatively 
low-temperature accretion disk of an extremely massive black 
hole \citep[e.g.][]{Laor2011}. While these explanations for 
WLQs have some attractive aspects, they cannot compellingly 
account for the more recently established notable \xray\ 
properties of WLQs described above---the shielding
model was designed to explain both the observed \xray\ 
and optical/UV properties of WLQs. Furthermore,
we note that the \citetalias{Luo2015} stacking-based 
measurements of \xray\ properties can only
constrain these properties on average, and it is possible 
that there are some intrinsically \xray\ weak objects among 
WLQs as well; i.e., quasars where the central engine simply
fails to generate \xrays\ at the nominal level. Such WLQs
would be similar to the well-studied intrinsically \xray\ 
weak quasar PHL~1811 \citep[e.g.][]{Leighly2007}, and the shielding model
would likely not be applicable for them.
\citetalias{Luo2015} provide further discussion on the possibility of some
intrinsically X-ray weak objects among the WLQs, though due to
relatively weak X-ray upper limits for many individual objects
this issue remains poorly understood.

\subsection{Motivation for the current study} \label{ssec-motivation}

Our initial investigations of the nature of WLQs via \xray\
observations primarily utilized brief exploratory \chandra\
observations of the most-extreme WLQs---we were
attempting to define the basic \xray\ nature of these
exceptional and poorly understood objects rather than observing well-defined, representative samples.
For example, we preferentially targeted WLQs with the weakest
possible C~{\sc iv} lines or those selected to have other 
exceptional properties, such as strong UV Fe~{\sc ii}/Fe~{\sc iii}
emission and large C~{\sc iv} blueshifts (see, e.g.
\citetalias{Luo2015} for details). While this approach was
successful for gaining the basic understanding described
above, it obviously has limitations. For example, the complex
underlying selection effects prevented proper assessment of 
if/how the fraction of \xray\ weak objects depends
upon C~{\sc iv} REW. Such a dependence would be of
considerable physical interest since, as described in Section~\ref{ssec_1_1},
the shielding model predicts a decline in this fraction with
rising C\,{\sc iv} REW---the predicted decline thus offers one
test of the shielding model. Indeed, if confirmed, the strength
and form of this reduction should provide insight into how the
covering factor of the shielding material regulates C\,{\sc iv}
line production.

Previously, we only had a small representative sample of WLQs with
\xray\ coverage for C~{\sc iv} REW values of $\lesssim 5$~\AA.
This sample was based on the WLQs originally identified by
\citet{Plotkin2010}, and it was completely disjoint from the
main population of typical quasars which mostly have C~{\sc iv} REW $=$ 15--100~\AA~(see Figure~\ref{civ}). We therefore
proposed in \chandra\ Cycle~17 to observe a sample of 14 quasars
with C~{\sc iv} REW values $\approx$ \hbox{5--15~\AA}, aiming to build
an unbiased \xray\ observational ``bridge'' between the \citet{Plotkin2010} WLQs and more typical quasars. With
these additional data in hand, one of our primary goals is
to perform assessments of the fraction of quasars that are \xray\
weak over the broad C~{\sc iv} REW range of \hbox{$\approx$ 3--100~\AA}.

In addition to studying the fraction of quasars that are
\xray\ weak as a function of C~{\sc iv} REW, we furthermore
desire to improve understanding of likely
connections between UV continuum/emission-line properties
and \xray\ weakness. \citetalias{Luo2015} reported that 
red UV continuum color [e.g. measured with $\Delta(g-i)$]
and large \hbox{2250--2650~\AA} Fe~{\sc ii} REW empirically
appear to serve as effective tracers of \xray\ weakness among
WLQs. These were not the tracers we had expected, and we are
still attempting to understand why they work as well as they
apparently do. Mild intrinsic reddening in the equatorial
plane \citep[e.g.][]{Elvis2012} and aspect-dependent accretion-disk
emission may be relevant; see \S5 and \S6 of \citetalias{Luo2015}.
However, the complex underlying selection effects and small
size of our current WLQ sample limit the analyses of such
tracers. The new \xray\ observations presented here should
aid in establishing which UV continuum/emission-line properties
are indeed most closely linked to \xray\ weakness, so that
interpretation of these connections can proceed with confidence.

\subsection{Paper layout and defined quantities}
The layout of this paper is as follows. 
We describe the sample selection and \hbox{X-ray} 
data analysis in Sections~\ref{sec-sp} and \ref{sec-Xray}. 
Multiwavelength analyses, including studies of 
\daox, UV emission-line properties, infrared 
to \hbox{X-ray} SEDs, and radio properties are given in 
Section~\ref{sec-multi}. The likely absorption nature of \hbox{X-ray} 
weakness is demonstrated in Section~\ref{sec-nature}. Changes
in the fraction of \xray\ weak quasars with high-ionization
emission-line strength are presented in Section~\ref{sec-frac},
and possible UV continuum/emission-line tracers of \hbox{X-ray}
weakness are discussed in Section~\ref{sec-UV}.

Throughout this paper, we use J2000 coordinates and a cosmology with $H_0=67.8$~km~s$^{-1}$~Mpc$^{-1}$, $\Omega_{\rm M}=0.308$, and $\Omega_{\Lambda}=0.692$ \citep{Planck2016}.

\section{SAMPLE SELECTION AND CHANDRA OBSERVATIONS} \label{sec-sp}
We have selected 32 representative WLQs with \iona{C}{iv} REW $\lesssim 15$ \AA~to investigate the nature of \xray\ weakness with reduced selection bias compared to our previous work. Fourteen of them were proposed and observed by \chandra~in Cycle 17, and the remaining 18 have archival \xray\ observations (see Sections \ref{ssec-select} and \ref{ssec-clean} for details). These 32 WLQs are divided into two subsamples, the ``bridge subsample'' and the ``extreme subsample'', according to their \iona{C}{iv} REW (see Section \ref{ssec-clean}), for study purposes. In addition, a larger sample of 63 WLQs, though not as consistently selected, is utilized to help probe potential tracers of \xray\ weakness (see Section \ref{ssec-full}). Two subsamples of the \citet{Gibson2008} Sample~B quasars are utilized (see Section \ref{ssec-clean}) as well for sample comparisons. All the samples we utilized for this paper are summarized in Table~\ref{summarytable}.

\subsection{Target selection and Chandra observations} \label{ssec-select}

Since typical quasars have \iona{C}{iv} REW $\approx$ 15--100 \AA~and WLQs with \iona{C}{iv} REW $\lesssim 5$ \AA~have been well-observed in the \xray\ band in recent years, our targeted \chandra~Cycle 17 WLQs have been selected from radio-quiet ($R < 10$) quasars in the SDSS Data Release~7 (DR7) quasar properties catalog \citep{Shen2011} with \hbox{\iona{C}{iv} REW = 5--15 \AA}~(which bridges the gap between previously \xray\ observed extreme WLQs and typical quasars). We also require $m_i$ $\leqslant$ 18.2, which is optically bright enough to allow economical snapshot \chandra~observations, and $1.7 < z < 2.2$, in order to have spectral coverage from \iona{Si}{iv} to \iona{Mg}{ii}. Objects with narrow absorption features around \iona{C}{iv}, BAL or mini-BAL features, or very red spectra with $\Delta(g-i) > 0.45$ were excluded since we want to avoid any potentially confusing \xray\ absorption associated with, e.g. quasar winds. 

Among the 20 selected quasars, four objects have already been targeted by \chandra~(SDSS J082508.75+115536.3, SDSS J132809.59+545452.7, and SDSS J153913.47+395423.4 were targeted in \chandra~Cycle~14; SDSS J215954.45--002150.1 was targeted in Cycle 11). Two further objects have been recorded in \xray\ survey catalogs: SDSS J113949.39+460012.9 in the full \xmm~slew-survey catalog \citep{Saxton2008} and SDSS J123326.03+451223.0 in the \textit{ROSAT} all-sky survey catalog \citep{Boller2016}. 
The relevant two \xray\ survey sources match the quasar positions to within (relatively large) positional uncertainties, and we do not find other likely optical counterparts for them. Thus, we take these two quasars to be \xray\ detected, although ideally we would prefer more secure identifications.
This left 14 objects for \chandra~observation. 

Our \chandra~observations were performed between 2015 Dec 28 and 2016 Aug 15 using the
Advanced CCD Imaging Spectrometer spectroscopic array (ACIS-S) \citep{Garmire2003} with VFAINT mode. 
The exposure times of our targets range from 3.1 to 4.8 ks. Table~\ref{xraytable} lists details of these 14 observations.

\subsection{Bridge WLQs vs. extreme WLQs in the representative sample} \label{ssec-clean}

As we have previously mentioned, WLQs with \iona{C}{iv} REW $\lesssim 5$~\AA~have been targeted in \hbox{X-rays} and studied in detail because of their extreme properties. 
 
In order to assemble an unbiased, statistically representative sample of WLQs with \iona{C}{iv} REW $\lesssim 5$ \AA, we select ten WLQs from the \citet{Plotkin2010} catalog, which identified WLQs from the full SDSS DR7 spectroscopic database as quasars having \iona{C}{iv} REW \hbox{$\lesssim 5$ \AA}. We used the same criteria as those in Section \ref{ssec-select} except for the $m_i$ and redshift requirements. An $i$-band magnitude of $m_i < 18.6$ is applied, and a redshift satisfying $1.5 < z < 2.5$ is required to have SDSS spectral coverage of the full \iona{C}{iv} region (which has been the major selector for weak emission lines) and the \hbox{2250--2650}~\AA~\iona{Fe}{ii} region (which has been proposed as a UV tracer of \xray\ weakness among WLQs). The differences among the selection criteria for $m_i$ and redshift are small, and should not affect the results.
All of the selected ten objects have archival \xray\ observations available.

Besides these ten objects from the \citet{Plotkin2010} catalog, we also include two additional WLQs with \xray\ observations: SDSS J094533.98+100950.1, identified by \citet{Hryniewicz2010}, and SDSS J090312.22+070832.4, identified by \citetalias{Wu2011}, satisfying the same criteria for radio loudness, \iona{C}{iv} REW, and redshift.

Together with the 20 WLQs with \iona{C}{iv} REW $=$ \hbox{5--15 \AA}~selected in Section \ref{ssec-select}, we have 32 (20+12) WLQs in total.
According to the ``shielding'' model of Section \ref{ssec_1_1}, we expect WLQs that are more heavily shielded (thus having weaker high-ionization UV emission-line strengths) as a result of thicker disks to be more inclined to be \xray\ weak.

Thus, we divided the 32 selected WLQs in the representative sample into two subsamples with equal numbers of objects according to their consistently re-measured \iona{C}{iv} REW (see Section \ref{sec-uvm}) in order to achieve the maximum statistical power in sample comparisons---this requirement leads to a subsample division threshold of 7~\AA.
Sixteen WLQs with \iona{C}{iv} REW $< 7.0$ \AA~are classified as the ``extreme subsample'', 
while 16 WLQs with \iona{C}{iv} REW = \hbox{7.0--15.5 \AA}~are classified as the ``bridge subsample''.
We only list archival X-ray observations of WLQs that have not yet been analyzed/published in Table~\ref{xraytable} along with Chandra Cycle 17 observations of targeted WLQs.

\begin{figure}
\begin{center}
\includegraphics[scale=0.46]{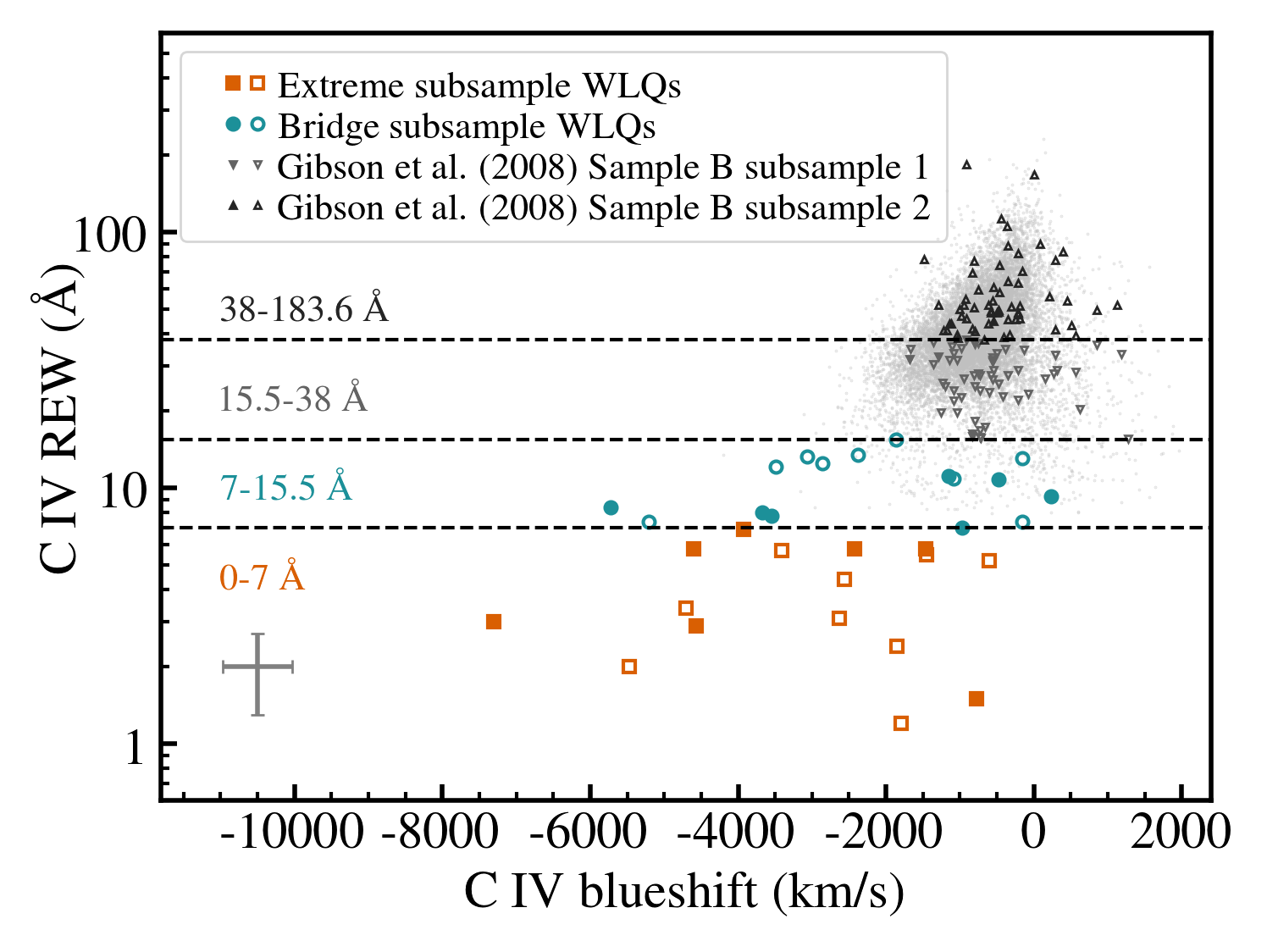}
\caption{ \iona{C}{iv} REW vs.\ \iona{C}{iv} blueshift. 
The extreme subsample, the bridge subsample, and two subsamples of the \citet{Gibson2008} Sample B quasars are distinguished using symbol types as labeled.
Solid symbols indicate \hbox{X-ray} weak WLQs, while open symbols indicate \xray\ normal WLQs.
The median uncertainties of {C}{ \sc iv} REW and {C}{ \sc iv} blueshift for WLQs are shown as the gray error bars in the bottom-left corner of the plot, and account for only the dominant error factors; i.e., the measurement error.
The dashed lines represent the \iona{C}{iv} REW division lines between the four subsamples, and the \iona{C}{iv} REW ranges of the four subsamples are labeled.
13582 typical radio-quiet quasars in \citet{Richards2011} are represented by the gray dots in the background.
}
\label{civ}
\end{center}
\end{figure}

Figure~\ref{civ} compares the \iona{C}{iv} REWs and blueshifts (as measured in Section \ref{sec-uvm})
of the objects in the extreme and bridge subsamples to those of typical radio-quiet quasars. The latter include quasars from \citet{Richards2011} and two subsamples of the \citet{Gibson2008} Sample~B quasars.\footnote{We utilized an ``ultra clean'' version of the \citet{Gibson2008} Sample~B of radio-quiet, non-BAL quasars, where seven likely BAL quasars and six likely mini-BAL quasars were removed as a result of the revision of the continuum model or the \iona{C}{iv} emission-line profile (see Footnote 16 of \citetalias{Wu2011}). Four objects with $\Delta(g-i) > 0.45$ (J1002+0203, J1245--0021, J1452+5804, and J1618+3456) were also removed (J1245--0021 and J1618+3456 are also likely mini-BAL quasars) to maintain consistency with the representative sample.} The \citet{Gibson2008} Sample~B quasars represent typical SDSS quasars that are radio-quiet ($R<10$) and lack BALs, which are also the selection requirements of our representative sample. 
All the quasars in the \citet{Gibson2008} Sample~B have quality \xray\ observations available, making comparisons of \xray\ properties with our representative sample possible. In this paper, two subsamples of the \citet{Gibson2008} Sample~B quasars are utilized; they are divided according to their \iona{C}{iv} REW, one set with 15.5 \AA~$<$ \iona{C}{iv} REW $<$ 38.0 \AA~ and the other with \iona{C}{iv} REW $\geqslant$ 38.0 \AA, with nearly equal size ($59/60$) to achieve the maximum statistical power in sample comparisons. They also have comparable redshift ranges ($1.7 < z < 2.7$) with our representative sample (see Table~\ref{summarytable} and Figure~\ref{zmag}).

Though selected only by \iona{C}{iv} REWs, we can also see the increasing trend in the \iona{C}{iv} blueshift values of the 32 selected WLQs when comparing with typical quasars or within the representative sample. The Eddington ratios of quasars generally increase as we move in Figure~\ref{civ} toward small \iona{C}{iv} REW and large \iona{C}{iv} blueshift \citep[e.g.][]{Richards2011,Shen2014,Sulentic2014,Sun2018}, and quasars accreting with high Eddington ratios are expected to have geometrically thick inner accretion disks \citep[e.g.][]{Abramowicz1988,Jiang2014,Jiang2017,Sadowski2014,Wang2014}. Thus, among our selected WLQs, we plausibly expect the presence of thick inner disks that can block ionizing photons from reaching the high-ionization BELR. These thick inner disks could also cause \xray\ weakness when oriented such that they block our line-of-sight to the compact \xray\ emission region (see Figure~\ref{schematic}).

\subsection{Full sample} \label{ssec-full}
In addition to performing analyses with the representative sample, we also utilize a larger ``full sample''. 
This sample is not as uniformly defined and unbiased as the representative sample, but its larger size makes it useful to improve the significance of our study and further assess the results obtained from the representative sample, especially when investigating UV tracers of \xray\ weakness.

There are 63 WLQs in our full sample. Besides the 32 WLQs in the representative sample, we also add 31 WLQs with \xray\ measurements that do not have BAL features or very red spectra: 19 from \citetalias{Luo2015}, 6 from \citetalias{Wu2011}, and 6 from \citetalias{Wu2012}. 

These quasars were originally targeted with rather complex selection rules for the \xray\ observations. Some of these objects have redshift values either too high or too low to allow spectral coverage of the full \iona{C}{iv} and \iona{Fe}{ii} regions. Some were targeted when studying ``PHL 1811 analogs''; additional requirements on \iona{Fe}{ii}/\iona{Fe}{iii} strength and \iona{C}{iv} blueshift rendered them not ideal objects to include in the representative sample for an unbiased study. 

\begin{figure}
\begin{center}
\includegraphics[scale=0.46]{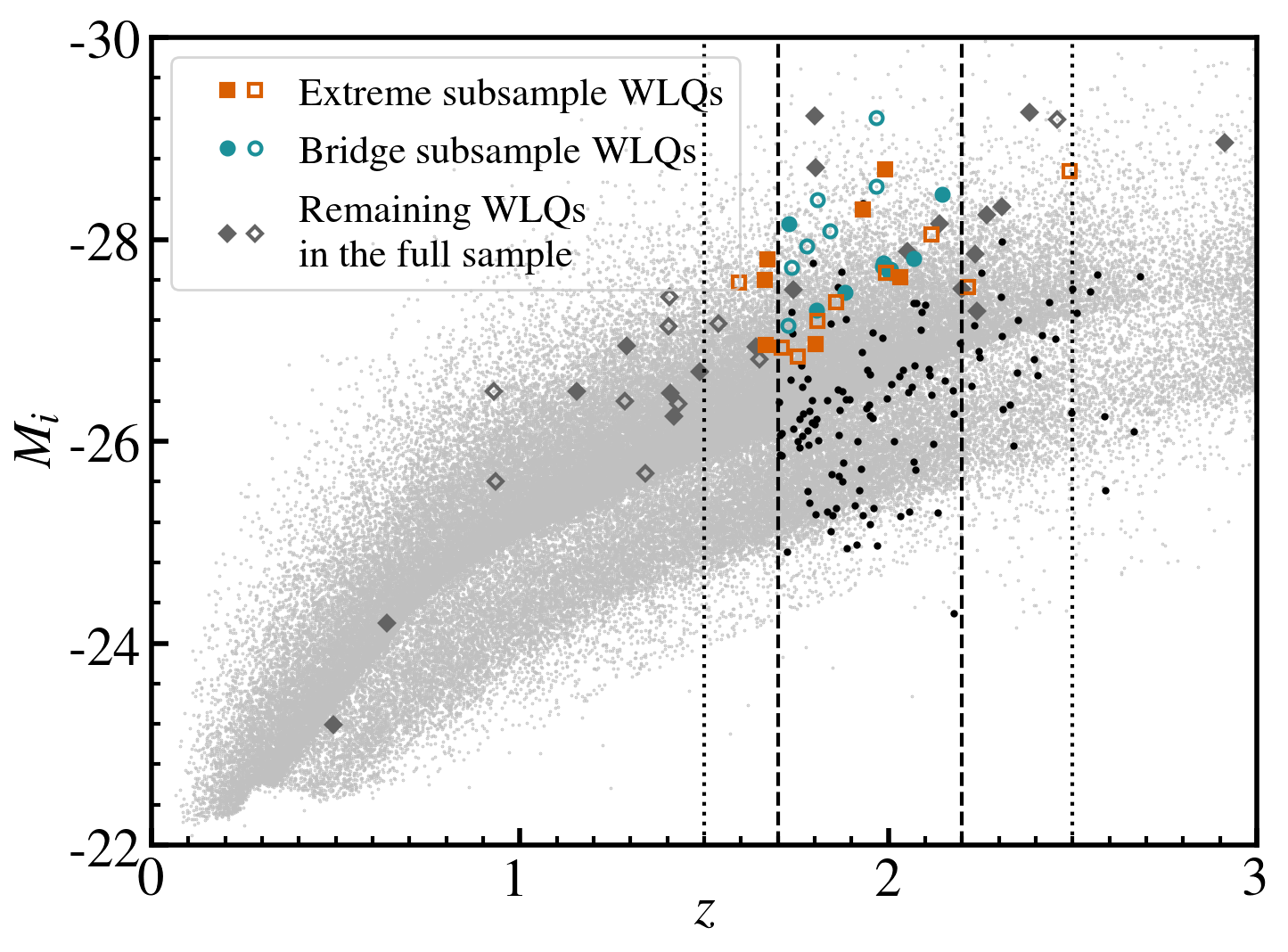}
\caption{Redshift vs.\ absolute $i$-band magnitude of the 63 WLQs in the full sample. 
Squares and circles represent WLQs in the representative sample,
and the diamonds indicate the remaining WLQs.
Solid symbols depict \xray\ weak WLQs, while
open symbols depict \xray\ normal WLQs.
The gray dots in the background indicate objects from the SDSS DR7 quasar catalog \citep{Schneider2010}, and the black dots in the background indicate quasars from \citet{Gibson2008} Sample B.  
The vertical dashed lines show the narrow redshift spread for the bridge subsample, 
and the vertical dotted lines show the redshift spread for the extreme subsample.}
\label{zmag}
\end{center}
\end{figure}

Figure~\ref{zmag} displays the redshift vs. absolute $i$-band magnitude plot of all the objects in the full sample. In comparison with the full sample, the representative sample is selected in a narrower redshift range.

\section{X-RAY PHOTOMETRIC ANALYSIS} \label{sec-Xray}
We processed the \chandra~Cycle 17 data using the Chandra Interactive Analysis of Observations (CIAO) tools \citep{Fruscione2006}.
The CHANDRA\_REPRO script was employed for the latest calibration.
The DEFLARE script was used to remove background flares with sigma clipping at a 3$\sigma$ level.
The final exposure times for each quasar are listed in Table~\ref{xraytable}.

We produced images for each source in the 0.5--2 keV (soft), 2--8 keV (hard), and 0.5--8 keV (full) bands from the cleaned event file using the standard \textit{ASCA} grade set (0, 2, 3, 4, 6).
WAVDETECT with a $\sqrt{2}$ sequence of wavelet scales (1, 1.414, 2, 2.828 and 4 pixels) was run in each band to
find \hbox{X-ray} sources with a false-positive probability threshold of 10$^{-6}$.

If we detect a source in no fewer than one band, the WAVDETECT position which is closest to the SDSS position is adopted as the source position. 
If WAVDETECT does not successfully identify any source, the SDSS position is adopted as the \hbox{X-ray} position.
We also ensure that no confusion has been made with source identifications (e.g. no other potential close sources).
The \hbox{X-ray-to-optical} offsets for the detected sources have a mean value of 0.54$''$, consistent with expectations.

Aperture photometry is performed in both the soft band and hard band, following the approach of \citetalias{Luo2015}.
For all the objects, we extracted source counts from a circular aperture with radius 2$''$ centered on the source position, 
which corresponds to encircled-energy fractions of 0.959 in the soft band and 0.907 in the hard band.
We extract background counts from source-free annular regions with an inner radius of 10$''$ and an outer radius of 40$''$.

In order to determine the significance of the source signal, a binomial no-source probability (the probability of detecting the source counts by chance when there is no real source), $P_{B}$ \citep[e.g.][]{Broos2007,Xue2011,Luo2013,Luo2015}, is computed for each band for each object; $P_{B}$ is defined as:
$$P_{B} (X\geqslant S) = \sum_{X=S}^{N} \frac{N!}{X!(N-X)!} p^{X} (1-p)^{N-X}$$

In this equation, $S$ represents the number of counts in the source area, $B$ represents the number of counts in the background area, and $N = S + B$.
$p=1/(1+BACKSCAL)$, where $BACKSCAL$ is the ratio of background-extraction area to source-extraction area.

When the calculated $P_{B}$ value is no more than 0.01, we consider the source to be detected in this band. This threshold is appropriate given that we are checking for detections only at pre-specified positions. The 1$\sigma$ errors of the source counts and background counts were obtained following \citet{Gehrels1986}. We followed Section 1.7.3 of \citet{Lyons1991} to calculate the 1$\sigma$ error of net counts.
When $P_{B} > 0.01$, the source is considered to be undetected in this band. The upper limits on the source counts for such cases were derived using the 90\% confidence-level table in \citet{Kraft1991}. The results from the above calculations are presented in Table~\ref{xraytable}.

The ratio of hard-band counts to soft-band counts is defined as the band ratio. The band ratio and its uncertainty (for sources detected in both bands) or upper limit (for sources only detected in the soft band) were obtained from the Bayesian Estimation of Hardness Ratios code BEHR \citep{Park2006}.
The 0.5--8 keV effective power-law photon index ($\Gamma_{\rm eff}$) or its lower limit for each source was derived from the band ratio using the CIAO script MODELFLUX, adopting a power-law spectrum with Galactic absorption. 
When sources are undetected in both the soft and hard bands, we are unable to place any constraints on $\Gamma_{\rm eff}$. 
Since luminous radio-quiet quasars typically have $\Gamma_{\rm eff} =1.8-2.0$ \citep[e.g.][]{Reeves1997,Just2007,Scott2011}, we adopt $\Gamma_{\rm eff} =1.9$ for the flux calculation of undetected sources. If such sources appear to be \xray\ weak (as defined in Section \ref{sec-aox}) after the first calculation, their \xray\ properties are re-calculated with $\Gamma_{\rm eff} =1.4$, which is the $\Gamma_{\rm eff}$ value from \xray\ stacking results for 30 \xray\ weak WLQs in \citetalias{Luo2015}.

\section{MULTIWAVELENGTH ANALYSES} \label{sec-multi}

\subsection{X-ray to optical properties} \label{sec-aox}

\begin{figure*}
\centering{
\includegraphics[scale=0.46]{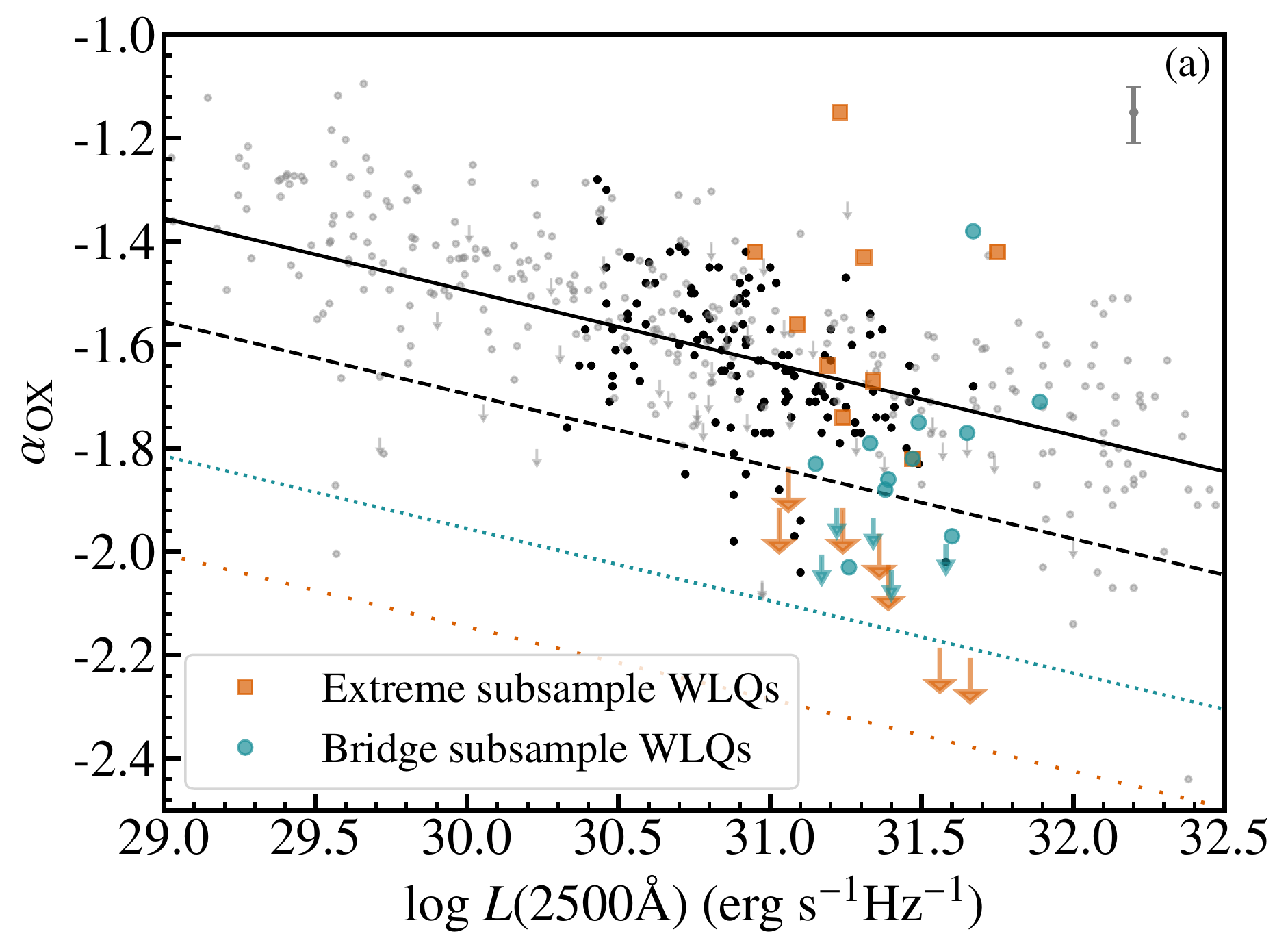}
\includegraphics[scale=0.46]{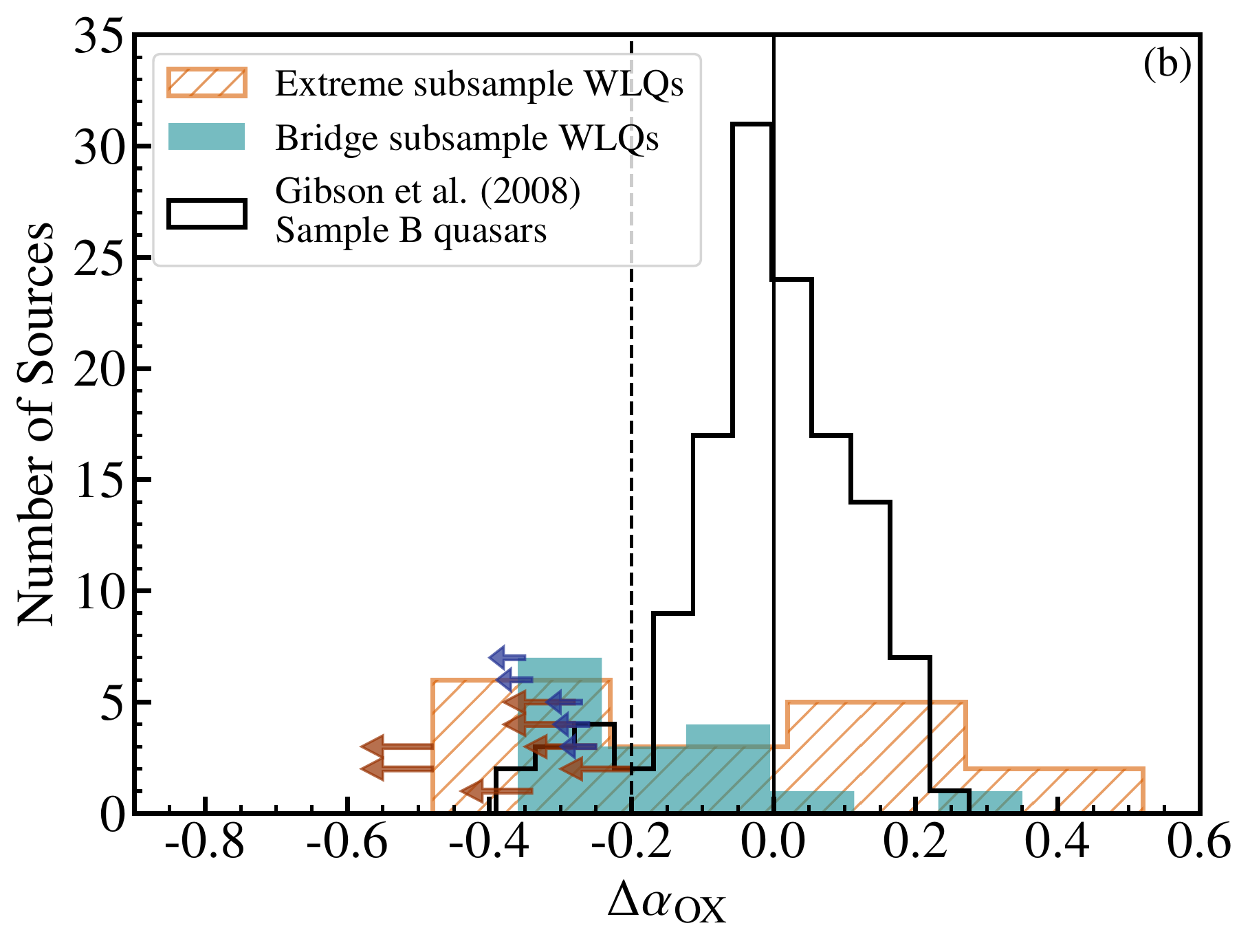}
}
\caption{(a). X-ray-to-optical power-law slope ($\alpha_{\rm OX}$) vs.\ 2500 \AA\
monochromatic luminosity for the bridge subsample WLQs (circles) and extreme subsample WLQs (squares).
In cases of X-ray non-detections, the 90\% confidence upper limits of $\alpha_{\rm OX}$ for extreme/bridge WLQs are represented by the downward long/short arrows.
For comparison, \citet{Gibson2008} Sample~B quasars are indicated by the black dots, and quasars from \citet{Steffen2006} and \citet{Just2007} are indicated by the gray dots and downward arrows, representing upper limits. The median measurement error of $\alpha_{\rm OX}$ (estimated from the count-rate uncertainty) for detected WLQs is represented by the gray error bar in the upper-right corner of the plot, while the measurement error of $L_{\rm 2500~{\textup{\AA}}}$ ($\approx 0.01~\rm dex$) is too small to be shown in the plot.
The solid line stands for the $\alpha_{\rm OX}$--$L_{\rm 2500~{\textup{\AA}}}$ relation from \citet{Just2007}; 
the dashed line ($\Delta\alpha_{\rm OX}=-0.2$) represents the adopted division between \xray\ normal and \xray\ weak quasars in this study.
The blue dotted line ($\Delta\alpha_{\rm OX}=-0.46$) marks the stacked mean of \xray\ weak WLQs in the bridge subsample, while the orange dotted line ($\Delta\alpha_{\rm OX}=-0.65$) marks the stacked upper limit of \xray\ weak WLQs in the extreme subsample (see Section \ref{ssec-stack} below).
(b). Distribution of $\Delta\alpha_{\rm OX}$ values. The solid histogram represents sources in the bridge subsample, while the 
hatched histogram represents sources in the extreme subsample.
The histogram bin widths are decided by Knuth's rule \citep{Knuth2006}.
The short and long arrows distinguish WLQs in the bridge and extreme subsamples for which the $\Delta\alpha_{\rm OX}$ value counted in the histogram is only an upper limit at the 90\% confidence level.
The unshaded histogram represents the $\Delta\alpha_{\rm OX}$ distribution for the \citet{Gibson2008} Sample B quasars, which is plotted for comparison.
The vertical dashed line (at $\Delta\alpha_{\rm OX}=-0.2$) represents the adopted threshold for defining \xray\ weakness in this study.
}
\label{alphaox}
\end{figure*}

We measured the \xray\ to optical power-law slope value ($\alpha_{\rm OX}$) for our \chandra~Cycle 17 sample objects and the archival objects that had not been previously analyzed, 
which was calculated from the rest frame 2500 \AA~and 2 keV flux densities (see Footnote \ref{aoxfootnote} for details).

The rest-frame 2500~\AA\ flux densities were taken from the \citet{Shen2011} SDSS DR7 quasar properties catalog.
By assuming a power-law spectrum modified by Galactic absorption (see Section \ref{sec-Xray} for the process of deriving $\Gamma_{\rm eff}$), we can derive the unabsorbed soft-band (0.5--2 keV) flux (which covers \hbox{2 keV} in the rest frame for all our sources) from the net count rate in the soft band using MODELFLUX, and thus calculate the rest-frame 2 keV flux densities. If we did not detect a source in the soft band, we could get an upper limit on $f_{2 {\rm keV}}$ following the same method using the upper limit for the soft-band net count rate. The flux densities and the calculated $\alpha_{\rm OX}$ values are listed in Table~\ref{aoxtable}.

One can derive an expected value of $\alpha_{\rm OX}$, $\alpha_{\rm OX}(L_{2500~\mathring{\rm{A}}})$,
from the empirical \hbox{$\alpha_{\rm OX}$--$L_{\rm 2500~{\textup{\AA}}}$} relation \citep[e.g.][]{Just2007} for typical optically-selected quasars.
The difference between the observed $\alpha_{\rm OX}$ and the expected value $\alpha_{\rm OX}(L_{2500~\mathring{\rm{A}}})$,
$\Delta\alpha_{\rm OX}=\alpha_{\rm OX}({\rm Observed})-\alpha_{\rm OX}(L_{2500~\mathring{\rm{A}}})$, 
provides a basic measurement of \xray\ weakness relative to the expected \xray\ luminosity and a factor of \xray\ weakness, $f_{\rm weak}$. 
Sources with $\Delta\alpha_{\rm OX} = -0.384$ have \xray\ fluxes $\approx 10\%$ of those of typical quasars, so we define $f_{\rm weak}$ as $10^{-\Delta\alpha_{\rm OX}/0.384}$,
which is approximately $403^{-\Delta\alpha_{\rm OX}}$.

The $\alpha_{\rm OX}$ vs.\ $L_{\rm 2500~{\textup{\AA}}}$ diagram
for our representative sample is presented in Figure~\ref{alphaox}(a).
Following \citetalias{Luo2015}, we use $\Delta\alpha_{\rm OX}=-0.2$ ($f_{\rm weak}=3.3$) to separate \xray\ weak quasars from \xray\ normal quasars, 
which corresponds to the $\approx$ 90\% single-sided lower confidence limit of the expected $\alpha_{\rm OX}$
distribution given the rms scatter in \citet{Steffen2006}.
A large fraction of our WLQs fall below this line, showing \xray\ weakness.
The distributions of $\Delta\alpha_{\rm OX}$ for the extreme subsample and bridge subsample are shown in Figure~\ref{alphaox}(b) and compared with the typical quasar distribution from the \citet{Gibson2008} Sample~B quasars.

\subsection{X-ray stacking analyses of WLQs} \label{ssec-stack}

The limited number of counts for each \xray\ weak WLQ has prevented us from carrying out individual \xray\ spectral analyses, but we could probe the average spectral properties of the population with stacking.
We performed the \xray\ stacking analyses by adding the extracted source and background counts of objects together and carried out the same \xray\ photometric analyses as in Section \ref{sec-Xray} for \xray\ weak WLQs in the extreme subsample, the bridge subsample, and the whole representative sample. The results of this exercise are given in Table~\ref{stacktable}.

X-ray weak WLQs in the whole representative sample have a stacked power-law effective photon index of $\Gamma_{\rm eff}=1.19_{-0.45}^{+0.56}$.
While $\Gamma_{\rm eff}$ could not be constrained for \xray\ weak WLQs in the extreme subsample due to the limited number of photons detected, the \xray\ weak WLQs in the bridge subsample have a stacked $\Gamma_{\rm eff}=1.09_{-0.45}^{+0.55}$. This result justifies our choice of $\Gamma_{\rm eff} = 1.4$ when performing \xray\ photometric analysis; i.e., \xray\ weak WLQs have harder $\Gamma_{\rm eff}$ compared with typical luminous radio-quiet quasars, which have $\Gamma\approx1.9$ (see Section \ref{sec-Xray}). 
This result also provides support to our shielding model by suggesting high levels of intrinsic \xray\ absorption, Compton reflection, and/or scattering, which are reflected by hard spectral shapes (see Section \ref{sec-nature} for further discussion).

Table~\ref{stacktable} also shows that the stacked $\Delta\alpha_{\rm OX}$ value of \xray\ weak WLQs is $-0.45 \pm0.11$ for the bridge sample (the error is calculated via bootstrapping source counts), and only an upper limit of $-0.64$ could be obtained for the extreme sample.

\subsection{UV emission-line measurements} \label{sec-uvm}
\begin{figure}
\begin{center}
\includegraphics[scale=0.46]{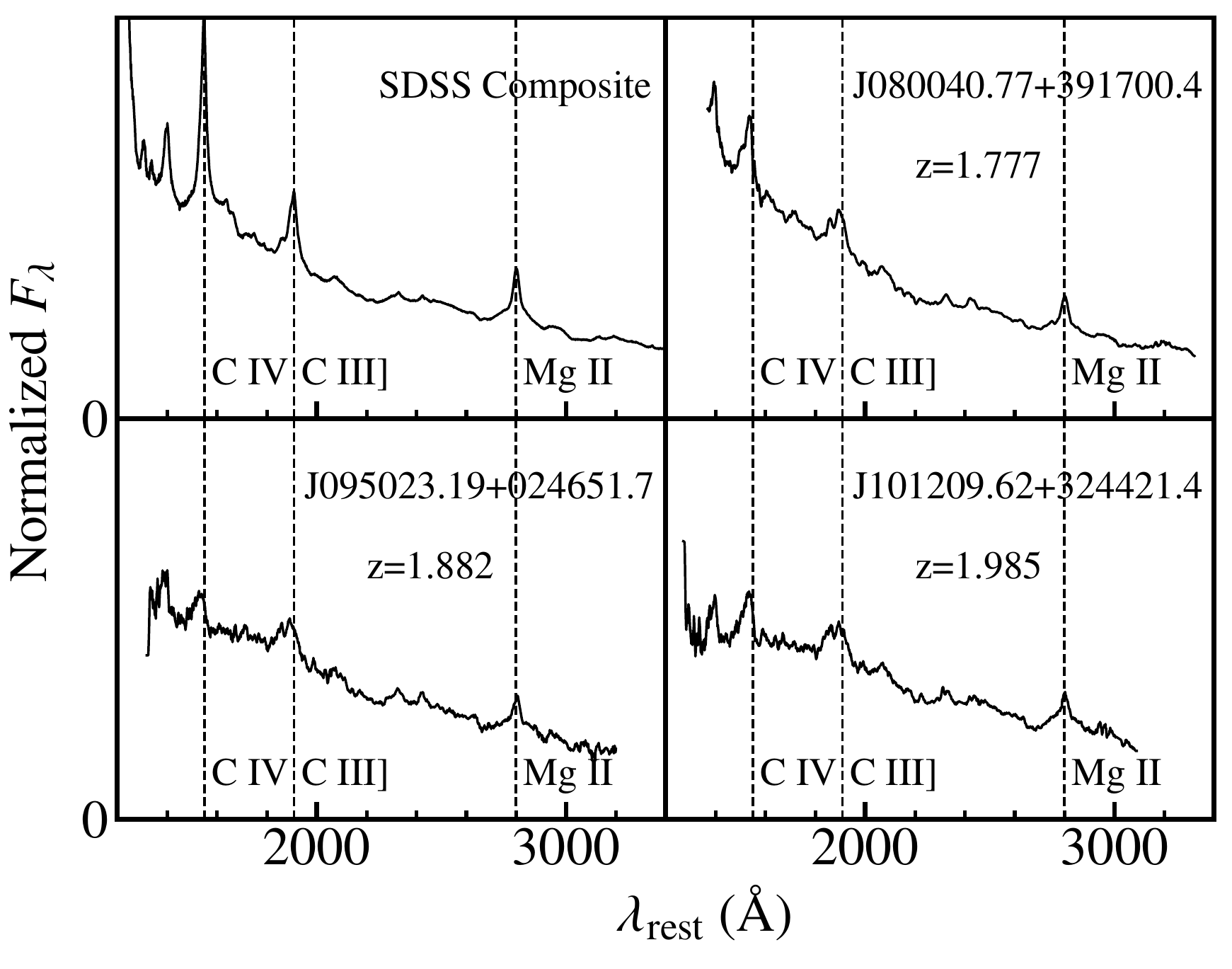}
\caption{SDSS spectra of WLQs in the bridge subsample.
A full version of this figure with all the spectra presented following the order of WLQs in Table~\ref{uvtable} is available in the online version.
The flux density in arbitrary linear units is plotted on the y-axis.
Dereddening and smoothing (with a 20-pixel boxcar) have been applied for each spectrum.
In the upper left panel, the composite spectrum of SDSS quasars from \citet{Vandenberk2001}
is presented for comparison.
}
\label{specall}
\end{center}
\end{figure}

Since some UV emission-line properties are considered to be potential tracers of \xray\ weakness, we performed emission-line measurements for the objects in the bridge subsample based on their SDSS spectra. The SDSS spectra of these WLQs are shown in Figure~\ref{specall}, compared with the SDSS quasar composite spectrum from \citet{Vandenberk2001}.
Most of the WLQs in the extreme subsample were previously measured in \citetalias{Luo2015}, and our \chandra~Cycle 17 targets making up the bridge subsample were analyzed in the same manner.

The UV emission-line properties listed in Table~\ref{uvtable} were measured following the methods in Section 2.2 of \citetalias{Wu2011} and Section 2.3 of \citetalias{Luo2015}, 
with the \citet{Hewett2010} redshift adopted.
The wavelength-fitting regions of the subtracted local power-law continuum for the \iona{Si}{iv}~$\lambda1397$, 
\iona{C}{iv}~$\lambda1549$, $\lambda1900$ complex
(mainly \iona{C}{iii}]~$\lambda1909$), 
\iona{Fe}{iii} UV48 $\lambda2080$,
and \iona{Mg}{ii}~$\lambda2799$ emission features are taken from Table~2 of \citet{Vandenberk2001}.
The UV \iona{Fe}{ii} was fitted between 2250~\AA\ and 2650~\AA, following \citetalias{Luo2015}.

\subsection{Infrared-to-X-ray SEDs} \label{ssec-sed}

\begin{figure*}
\centerline{
\includegraphics[scale=0.46]{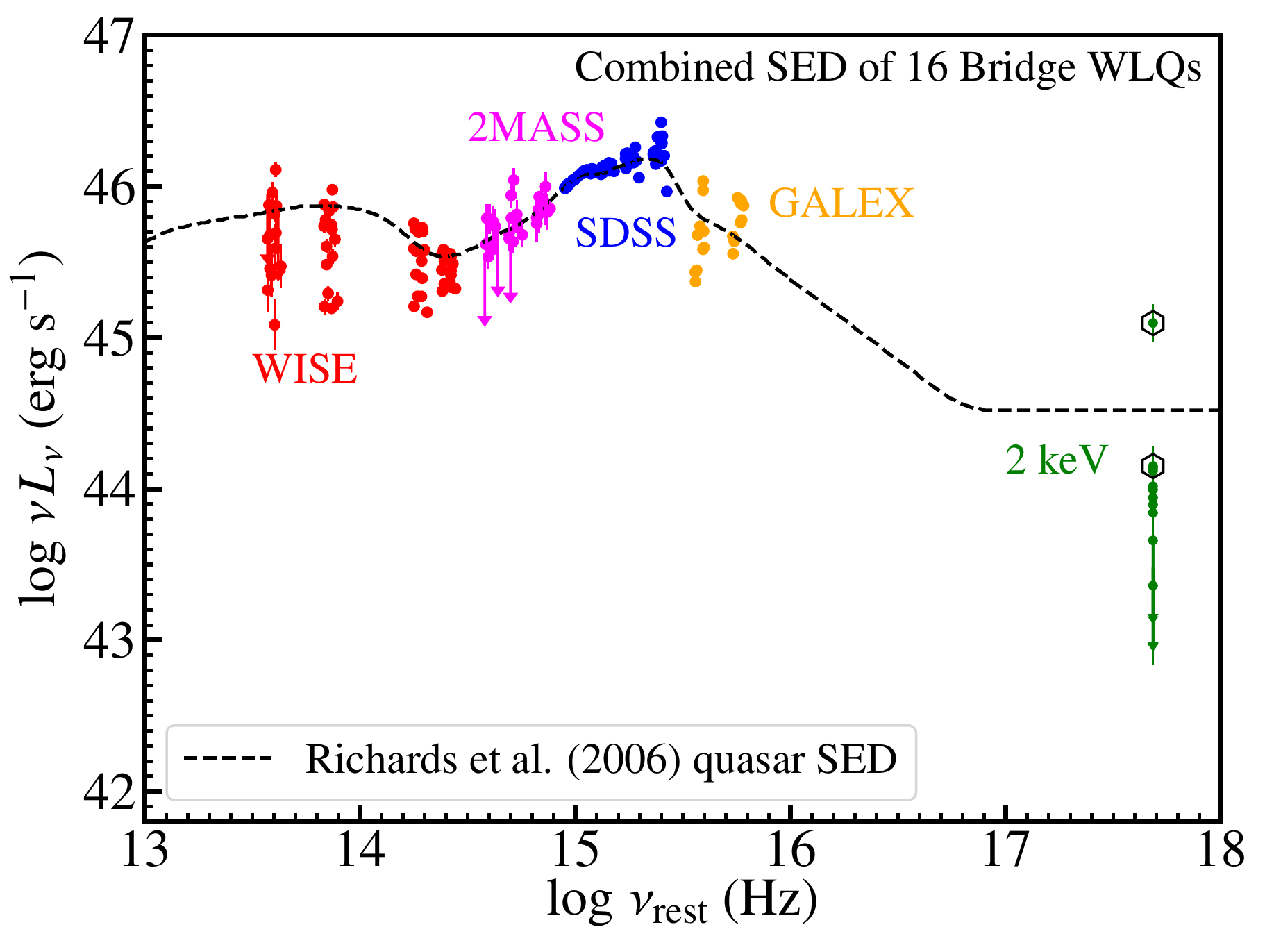}
\includegraphics[scale=0.46]{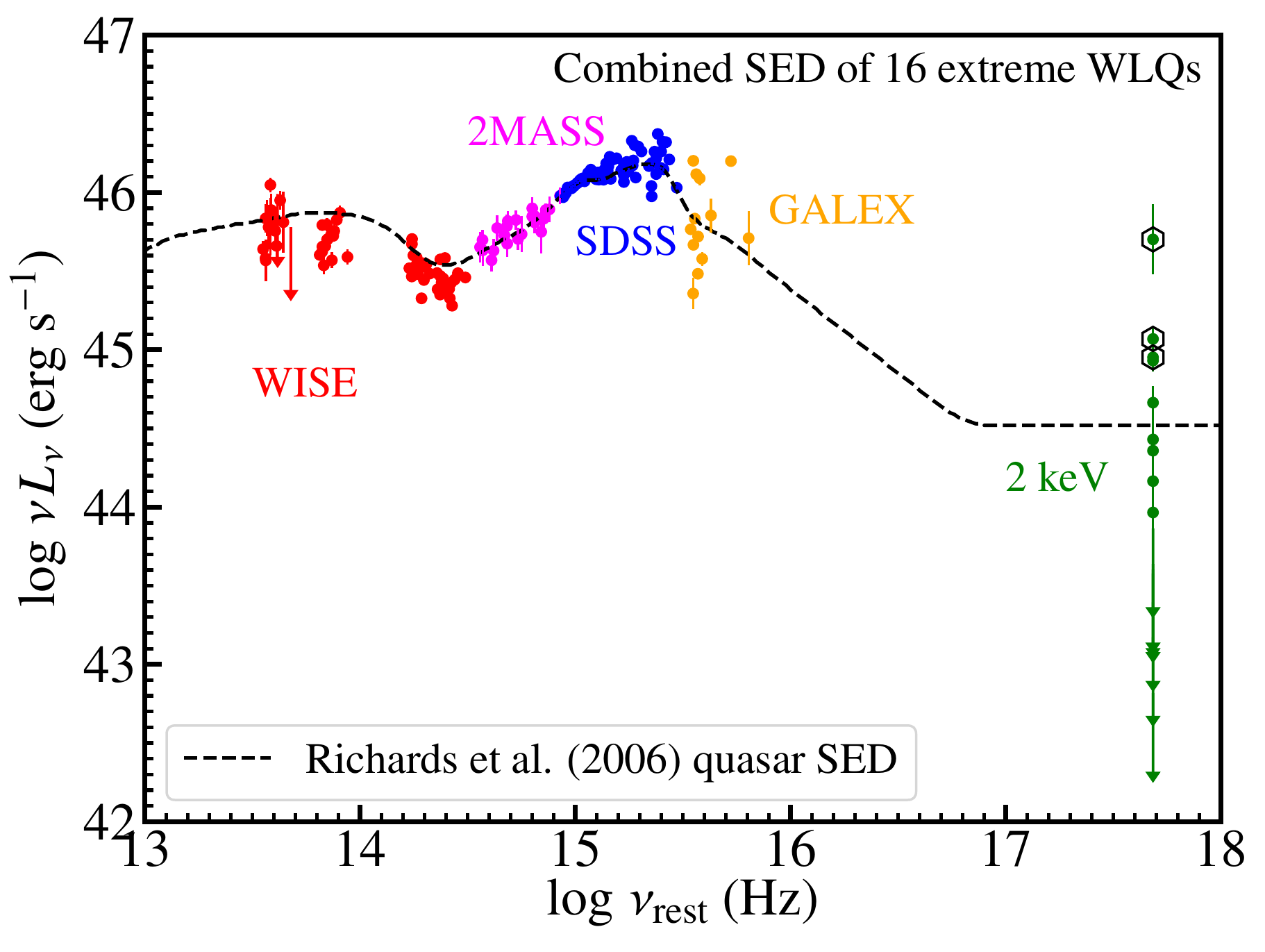}
}
\caption{Combined SEDs of 16 WLQs in the bridge subsample (left) and 16 WLQs in the extreme subsample (right).
The IR-to-UV SED data obtained from the {\it WISE}, 2MASS, SDSS, and {\it GALEX} catalogs are represented by red, magenta, blue, and orange points individually 
(the arrows represent 95\% confidence upper limits for {\it WISE} and 2MASS data).
The green data points and arrows represent luminosities and 90\% confidence upper limits at 2~keV, and those with black hexagons indicate sources that were detected by FIRST.
The SED of each WLQ was scaled at rest-frame 3000~\AA\ ($10^{15}$ Hz) to the composite SED (dashed line) of optically
luminous quasars from \citet{Richards2006}.
(In the combined SEDs, we removed the data points collected from {\it GALEX} that are influenced by the
Lyman break).
}
\label{sed}
\end{figure*}

The infrared (IR) to X-ray continuum SEDs for the bridge WLQ and the extreme WLQ subsamples were constructed with photometric data from the
 {\it Wide-field Infrared Survey Explorer} ({\it WISE}; \citealt{wright2010}), Two Micron All Sky Survey (2MASS; \citealt{Skrutskie2006}), SDSS, {\it Galaxy Evolution Explorer} ({\it GALEX}; \citealt{Martin2005}), and \chandra.
 
\textit{WISE} measured our targets at 3.4, 4.6, 12, and 22 $\mu \rm m$.
We convert \textit{WISE} magnitudes of our sources to flux densities following the method of \citet{wright2010}.
2MASS measured targets in the $J, H$, and $K_s$ bands, and we convert the magnitudes to flux densities following \citet{Cohen2003}.
\textit{GALEX} measured AB magnitudes in the FUV band \hbox{(1350--1750 \AA)} and NUV band \hbox{(1750--2750 \AA)}. We applied the dereddening method in \citet{Cardelli1989} and \citet{Odonnell1994} to correct for the Galactic extinction.

Figure~\ref{sed} displays the combined SEDs of the 16 WLQs in the bridge subsample and the 16 WLQs in the extreme subsample.
The composite SED of optically luminous SDSS quasars from \citet{Richards2006} is included for comparison. 
All the WLQs in the representative sample have IR-to-UV SEDs similar to those of typical quasars, which further eliminates the possibility of contamination from BL Lacs \citep[e.g.][]{Lane2011}. This similarity also justifies that our calculation of $\Delta\alpha_{\rm OX}$, the observed deviation from the expectation value of the \xray\ to optical power-law slope, is valid for probing the degree of \xray\ weakness, since it is not affected by any anomaly in optical properties.

\subsection{Radio properties} \label{sec-radio}
 
We only include radio-quiet ($R<10$) quasars in samples used in this study.
We computed the radio-loudness parameter as $R=f_{5~{\rm GHz}}/f_{\rm 4400~{\textup{\AA}}}$ \citep[e.g.][]{Kellermann1989}. 

The rest-frame 5 GHz flux densities $f_{5~{\rm GHz}}$ 
were obtained by checking the Faint Images of the Radio Sky
at Twenty-centimeters (FIRST) survey catalog \citep{Becker1995,White1997} at the 
source position and converting the measured 1.4 GHz result to rest-frame 5 GHz with an assumed radio power-law index $\rm \alpha_r = -0.5$ ($f_\nu \propto \nu^{\rm \alpha_r}$; \citealt{Kellermann1989}).
If sources were not detected by the FIRST survey, we obtained upper limits
on the radio fluxes as $0.25+3\sigma_{\rm rms}$~mJy,
where 0.25 mJy stands for the CLEAN bias \citep{White1997} and 
$\sigma_{\rm rms}$ represents the rms noise at the
source position in the FIRST survey.
The rest-frame 4400 \AA~flux densities were converted from the rest-frame 2500 \AA~flux densities,
assuming an optical power-law slope of $\rm \alpha_o = -0.5$ ($f_\nu \propto \nu^{\rm \alpha_o}$; e.g. \citealt{Richstone1980,Vandenberk2001}).
The calculated $R$ values for bridge WLQs are listed in the last column of Table~\ref{aoxtable}.
 
Among the 16 extreme WLQs, SDSS J084424.24+124546.5, SDSS J113949.39+460012.9, and SDSS J115637.02+184856.5
have point-like radio sources detected at the source positions. This is also the case for SDSS J123326.03+451223.0 and SDSS J215954.45--002150.1 among the 16 bridge WLQs. In all cases, these 5 objects are still radio quiet with $R=2.8-8.0$, and a mean $R$ of 5.0. We cannot completely exclude the possibility of some jet-linked \xray\ contamination for these five objects \citep[e.g.][]{Miller2011}, and they all proved to be \xray\ normal WLQs with relatively high $\Delta\alpha_{\rm OX}$. This behavior can also be seen clearly in Figure~\ref{sed}
 where all these objects with radio detections (noted with black hexagons) show rest-frame 2 keV luminosities above the typical luminosities.

We have performed the same investigations as those on all the WLQs in the representative sample (see Section \ref{sec-r}) for WLQs with no radio source detected, and we verify that our results are not materially influenced by including sources with radio detections. Thus, these WLQs are retained in the representative sample.
 
\section{RESULTS AND DISCUSSION} \label{sec-r}

\subsection{The nature of X-ray weakness} \label{sec-nature}
In principle, a quasar can appear \xray\ weak due to 
either \xray\ absorption or intrinsic \xray\ weakness
(see Section~\ref{ssec_1_1} for further discussion). One important
tool for discriminating between these two scenarios
involves examination of \xray\ spectral properties; e.g. 
a hard \xray\ continuum shape, especially at high 
rest-frame energies, is generally indicative of a spectrum
dominated by absorption, Compton reflection, and/or 
scattering. In this section, we therefore examine the 
\xray\ spectral properties of our WLQ samples to the 
extent the data allow. 

Power-law effective photon indices ($\Gamma_{\rm eff}$) derived from band ratios are used to assess the \xray\ continuum shape. \citetalias{Wu2011}, \citetalias{Wu2012}, and \citetalias{Luo2015} all found relatively flat/hard effective power-law photon indices for their \xray\ weak objects, suggesting heavy absorption.

The \xray\ weak WLQs in our representative sample have a mean redshift of 1.884 (see Table~\ref{stacktable}), which is useful for probing the level 
of absorption since the \xray\ fluxes in the rest frame are at relatively high energies ($\approx 2-23$~keV). 
However, only 1 out of 7 \xray\ weak WLQs in the bridge subsample has enough detected counts
to constrain the $\Gamma_{\rm eff}$, which is $0.9_{-0.4}^{+0.5}$ (see Table~\ref{xraytable}), and no \xray\ weak
 WLQs in the extreme subsample has constrained $\Gamma_{\rm eff}$ (\citetalias{Wu2011}, \citetalias{Wu2012}, \citetalias{Luo2015}).
Thus, we must rely on the results from \hbox{X-ray} stacking analyses (see Section \ref{ssec-stack}) to probe the nature of \xray\ weakness.

As shown in Table~\ref{stacktable}, all 14 \xray\ weak WLQs in the representative sample have a stacked $\Gamma_{\rm eff}=1.19_{-0.45}^{+0.56}$, and the 7 \xray\ weak WLQs in the bridge subsample have a stacked $\Gamma_{\rm eff}=1.09_{-0.45}^{+0.55}$.
These numbers are consistent with the shielding model by showing the flatness of the spectrum compared with
$\Gamma\approx1.9$ expected in the case of typical luminous radio-quiet quasars (see Section \ref{sec-Xray}). If an intrinsic $\Gamma \approx 1.9$ is assumed, $N_{\rm H} \gtrsim 10^{23}~\rm cm^{-2}$ is required to produce these stacked $\Gamma_{\rm eff}$ values at rest-frame $2-23$ keV.

We caution that we cannot actually constrain the stacked $\Gamma_{\rm eff}$ for the extreme subsample, and the stacking results for the bridge subsample and the representative sample as a whole might be dominated by only a few sources. Thus, we cannot fully exclude the possibility that there might be some intrinsically \xray\ weak quasars within our sample (also see Section~1.1). Nevertheless, the absorption scenario is still statistically preferred, which is consistent with the thick inner-disk model.

\subsection{The fraction of X-ray weakness} \label{sec-frac}

\begin{figure*}
\centerline{
\includegraphics[scale=0.46]{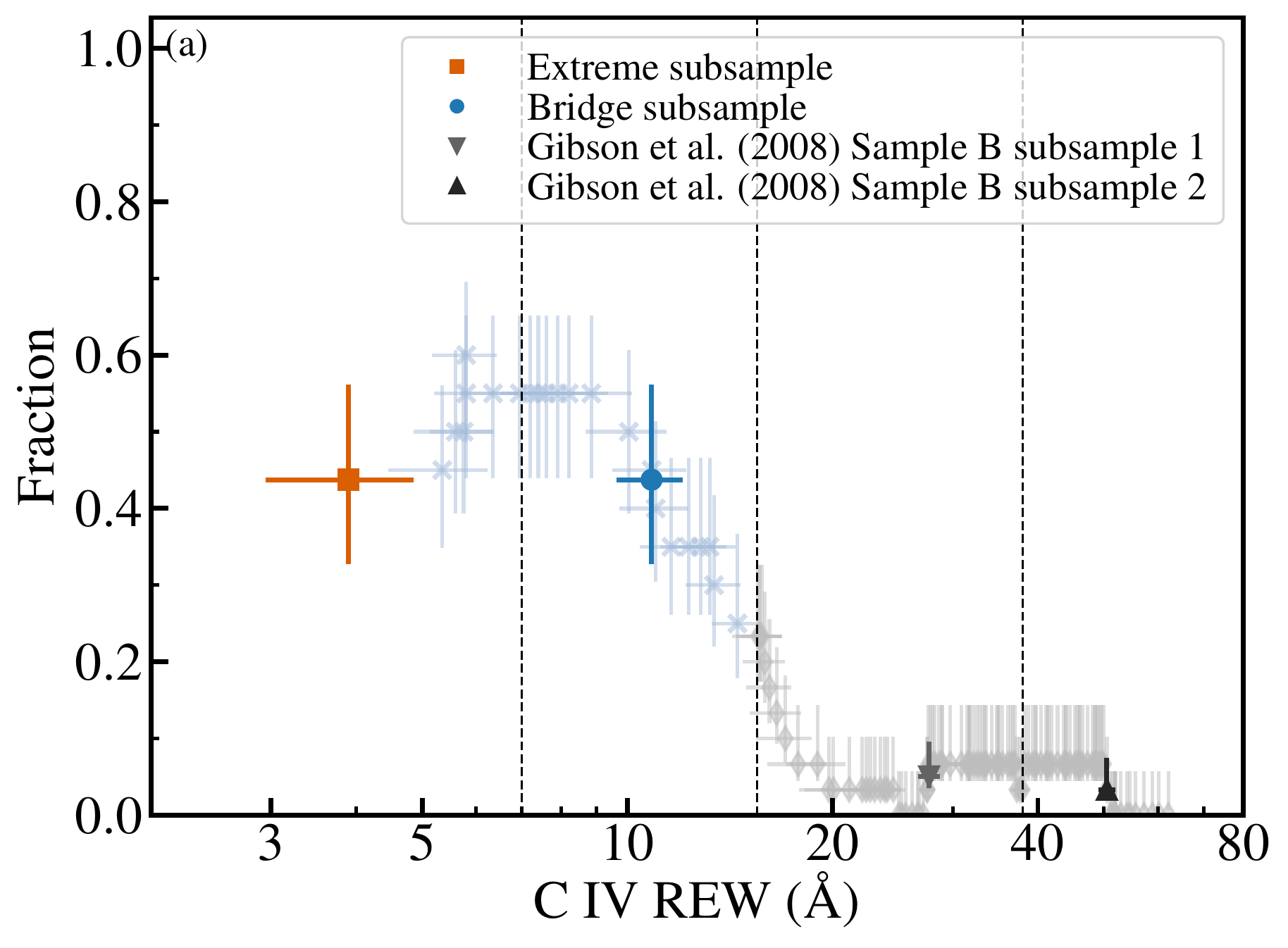}
\includegraphics[scale=0.46]{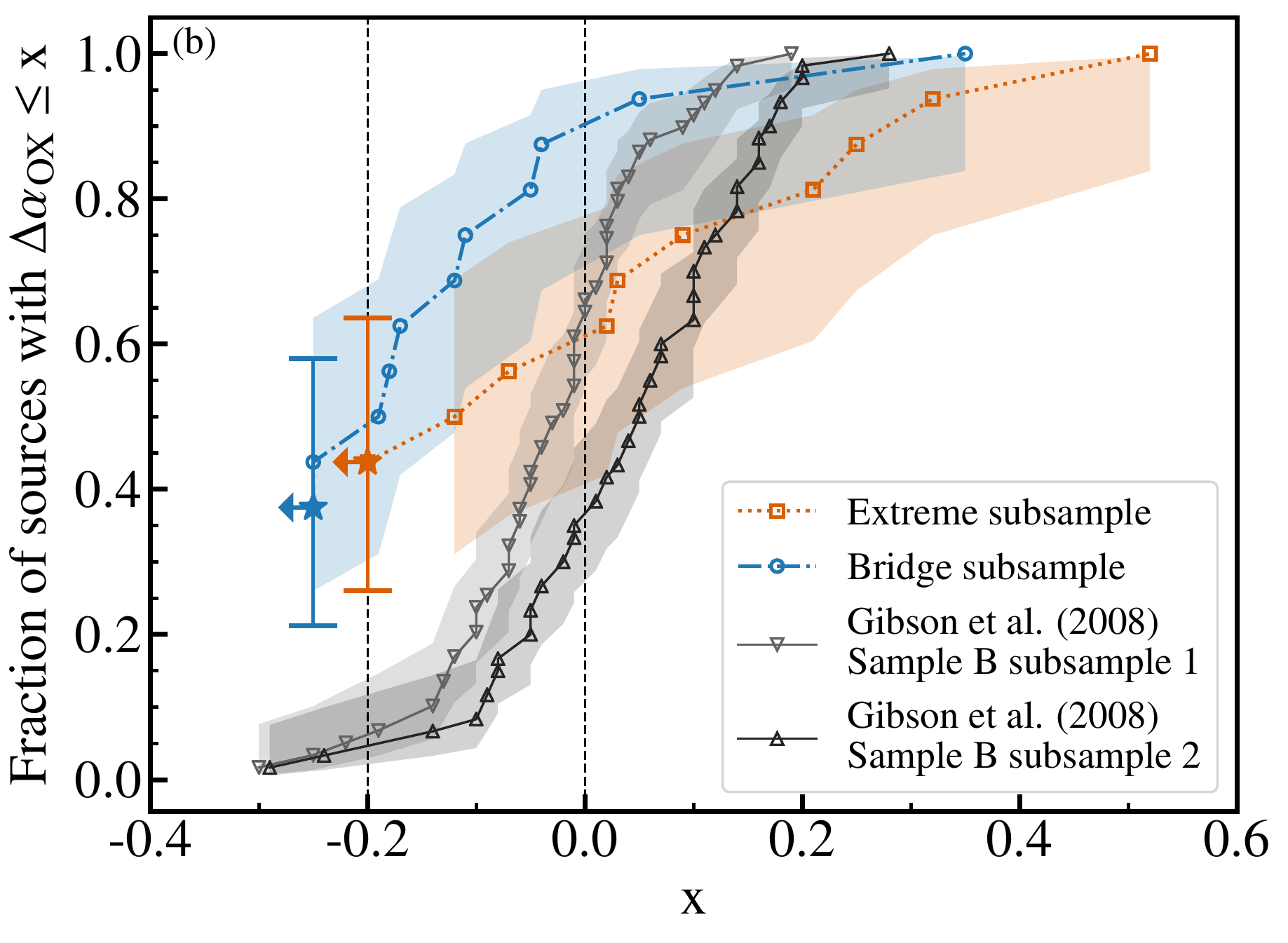}
}
\caption{(a). The fraction of X-ray weak quasars vs. the \iona{C}{iv} REW for the extreme subsample, bridge subsample, and two subsamples of the \citet{Gibson2008} Sample B quasars. 
The median \iona{C}{iv} REW for each subsample is plotted with error bars calculated from bootstrap.
The \xray\ weak fraction for each subsample is displayed with 1$\sigma$ error bars calculated from \citet{Cameron2011}.
The vertical dashed lines mark the divisions of \iona{C}{iv} REW between the four subsamples. 
The underlying lightly shaded crosses (thin diamonds) and their error bars represent the running-window results with a window size of 20 (30) objects using the four subsamples together.
(b). The fraction of sources having $\Delta\alpha_{\rm OX} \leqslant x$ vs.\ $x$
for the extreme subsample, bridge subsample, and two subsamples of the \citet{Gibson2008} Sample B quasars. 
For $\Delta\alpha_{\rm OX}$ below the measured value closest to the position of the star (which is the highest upper limit of $\Delta\alpha_{\rm OX}$ we obtained for individual sources in the subsample), we could not constrain the relevant fraction. 
The shaded regions (and the error bars attached to the stars) represent the 90\% confidence intervals for the fractions, calculated following \citet{Cameron2011}.
}
\label{fraction}
\end{figure*}

Nearly half of the quasars (43.8\%) in the whole representative WLQ sample 
are \hbox{X-ray} weak, which is a much larger fraction than reported for typical radio-quiet quasars in \citet{Gibson2008} 
Sample~B (7.6\%).
Among the 16 bridge-subsample WLQs, 7 are \hbox{X-ray} 
weak, and the \hbox{X-ray} weak fraction is also 43.8\%. 
For the extreme subsample, the number of \xray\ weak quasars and the \xray\ weak fraction are the same (see Table~\ref{fractable}). 

Assuming the binomial distribution, we 
use Table~1 of \citet{Cameron2011} to get the $1\sigma$ confidence 
intervals on the \hbox{X-ray} weak fractions, which are \hbox{32.7--56.2\%} 
for both the extreme and bridge subsamples.
If sources with radio detections are removed from the 
representative sample, the \hbox{X-ray} weak fraction is 7/14 
(50.0$^{+12.5}_{-12.5}$\%) among the bridge-subsample WLQs 
and 7/13 (53.8$^{+12.4}_{-13.4}$\%) among the extreme-subsample WLQs. 

In order to investigate the dependence of the \hbox{X-ray} weak 
fraction upon the \iona{C}{iv} REW more broadly, we have utilized the \citet{Gibson2008} Sample~B quasars.
We have used Fisher's exact test (a 
contingency table test valid for all sample sizes) to compare 
the fractions of \hbox{X-ray} weak quasars among the extreme-subsample WLQs, 
the bridge-subsample WLQs, and the two \citet{Gibson2008} subsamples. Table~\ref{fractable} 
lists, for all subsample pairs, the results from Fisher's exact 
test assessing the null hypothesis that the two subsamples have 
an equal fraction of \hbox{X-ray} weak quasars. Not
all pairwise comparisons yield significant results. However, there
is overall a highly significant trend showing a decline in the 
fraction of \hbox{X-ray} weak quasars with rising \iona{C}{iv} REW.
There is also an apparent trend of increasing median
$\Delta\alpha_{\rm OX}$ with rising \iona{C}{iv} REW.

Figure~\ref{fraction}(a) displays the \hbox{X-ray} weak 
fractions of the extreme and bridge subsamples of WLQs, together 
with those for the two \citet{Gibson2008} subsamples. 
The fraction of sources with $\Delta\alpha_{\rm OX} \leqslant x$ for each of the four subsamples can be seen in Figure~\ref{fraction}(b).
The clear systematic decline in the \hbox{X-ray} weak fraction with 
increasing \iona{C}{iv} REW is apparent. In the lightly shaded
background of Figure~\ref{fraction}(a), we also provide 
``running window'' results. To derive these values, we combine the quasars in 
the four subsamples together, sort them by their \iona{C}{iv} REW values, 
construct sets of quasars falling within a running window (which is
stepped across the quasars with a step size of one quasar), and 
calculate the \hbox{X-ray} weak fraction and its $1\sigma$ 
confidence intervals for each window (using the same methodology
as was used for the four subsamples). The adopted width of the 
running window is 20 quasars for \iona{C}{iv} REW $<$ 15~\AA\ 
and 30 quasars for higher \iona{C}{iv} REW. The running-window 
results also clearly show the systematic drop in the \hbox{X-ray} 
weak fraction with increasing \iona{C}{iv} REW, providing a 
more continuous view of this trend. The drop is especially pronounced 
in the \hbox{10--20 \AA} range. Due to the limited sample size of WLQs,
we cannot observe any statistically significant changes at {C}{ \sc iv} REW $<$ 10~\AA, 
which also prevents statistically significant differentiation of the extreme and bridge subsample areas in Figure~\ref{fraction}(b).

Although still limited by the sample sizes at low \iona{C}{iv} REW, 
we have now established using a relatively well-defined sample the overall trend that quasars with weaker 
\iona{C}{iv} are more likely to be \hbox{X-ray} weak. This trend is
strong, with the X-ray weak fraction declining by a factor of 
$\approx 13$ over the range of \iona{C}{iv} REW that we probe. 
The basic behavior is what is expected 
from the thick inner-disk model (see Figure~\ref{schematic}) put forward by \citetalias{Luo2015}.
Smaller \iona{C}{iv} REWs are associated with 
a thicker inner accretion disk that more effectively blocks
ionizing photons from reaching the high-ionization BELR (likely
due to higher Eddington ratios in such systems; e.g.
\citealt{Richards2011,Shen2014,Sulentic2014,Sun2018}). 
For random quasar orientations on the sky, when the \iona{C}{iv} REW
is small the thicker inner accretion disk is also more likely 
to block our line-of-sight to the central \hbox{X-ray} source, 
leading to a higher fraction of \hbox{X-ray} weakness. 

It is challenging to associate the drop in the \xray\ weak
fraction at \iona{C}{iv} REW values of 10--20~\AA\ with 
a specific range of the Eddington ratio, but here we provide
some basic estimates. For example, if we consider the 
anti-correlation found between $\rm H\beta$-based $L/L_{\rm Edd}$ 
and \iona{C}{iv} REW \citep[e.g.][]{SL2015}, the inferred 
$L/L_{\rm Edd}$ is $\approx 1$ at 20~\AA\ and 
$\approx 4$ at 10~\AA. In such a rapid-accretion regime, the 
inner accretion flow indeed likely becomes geometrically 
thick, and this $L/L_{\rm Edd}$ range is consistent
with other estimates (see Section~5.3 of \citetalias{Luo2015}).  
This being said, the \hbox{$L/L_{\rm Edd}\approx 1$--4} 
range of the $L/L_{\rm Edd}$ vs.\ \iona{C}{iv} REW
anti-correlation is not well sampled by observation and
is thus uncertain, and $\rm H\beta$-based estimates of 
black-hole masses may become unreliable for such rapidly 
accreting systems \citep[e.g.][]{Marconi2009}. Thus, the 
estimated $L/L_{\rm Edd}$ values associated with the drop 
in \xray\ weak fraction should be considered with caution. 

Considering the model in Figure~\ref{schematic}, the lack of clear dependence
of the \xray\ weak fraction upon \iona{C}{iv} REW for 
\iona{C}{iv} REW values of \hbox{$\approx 4$--10~\AA}
is somewhat surprising. Based on this range 
in \iona{C}{iv} REW of a factor of $\approx 2.5$, in a
shielding model one might have expected a rise by
a similar factor in the \xray\ weak fraction toward 
small \iona{C}{iv} REW values. However, we do not 
consider this discrepancy highly problematic at present
given the small sample sizes and the fact that other
factors will also influence \iona{C}{iv} REW values
(e.g. gas metallicity and anisotropic line emission; 
see Section \ref{ssec_1_1}). Further observations are required
for proper assessment of this issue. 

\subsection{Spectral tracers of X-ray weakness} \label{sec-UV}

Previous studies \citep[e.g.][]{Gibson2008,Luo2015} have demonstrated that various UV emission-line and continuum properties may be tracers of \xray\ weakness among quasars.
Nonparametric correlation tests are an effective probe of potential links between UV properties and $\Delta\alpha_{\rm OX}$. 
The Spearman rank-order test in the Astronomical Survival Analysis package (ASURV; e.g. \citealt{Feigelson1985,Lavalley1992}) was performed to check whether such correlations exist for several UV emission-line/continuum properties including the REW, blueshift, and FWHM of \iona{C}{iv};  
REWs of the \iona{Si}{iv}, $\lambda1900$, \iona{Fe}{ii}, \iona{Fe}{iii}, and \iona{Mg}{ii} emission features; 
and the relative SDSS color $\Delta(g-i)$. 
We have also applied the Peto-Prentice test \citep{Prentice1979} in the ASURV package \citep[e.g.][]{Feigelson1985,Lavalley1992} to assess whether \xray\ weak WLQs differ from \xray\ normal WLQs in their UV properties.
The results are presented in Table~\ref{pptable}. 
All the above statistical tests were done for both the representative sample and the full sample.
In addition, we have verified that the inclusion of sources with radio detections does not affect the results significantly. 

\subsubsection{Relative SDSS color $\Delta(g-i)$}
The SDSS color $g-i$ has a redshift dependence \citep[e.g.][]{Richards2001}, since the corresponding rest-frame wavelength ranges of the $g$ band and $i$ band change with redshift.
Typical rest-frame wavelengths being covered in our study (with a median redshift of $z\approx 1.9$) by the $g$ band and $i$ band are $1300-1900$~\AA~and $2300-2900$~\AA, respectively. 
In order to require the relative SDSS color $\Delta(g-i)$\footnote{$\Delta(g-i)=(g-i)-\langle(g-i)\rangle_{\rm redshift}$, where $\langle(g-i)\rangle_{\rm redshift}$ is the median $g-i$ color of SDSS quasars at a given redshift (see \citealt{Richards2003} for details).} to reflect UV (rather than optical) properties, we only utilize WLQs with $z>1$ when carrying out statistical tests of $\Delta(g-i)$ in the case of the full sample. 

\begin{figure*}
\centerline{
\includegraphics[scale=0.46]{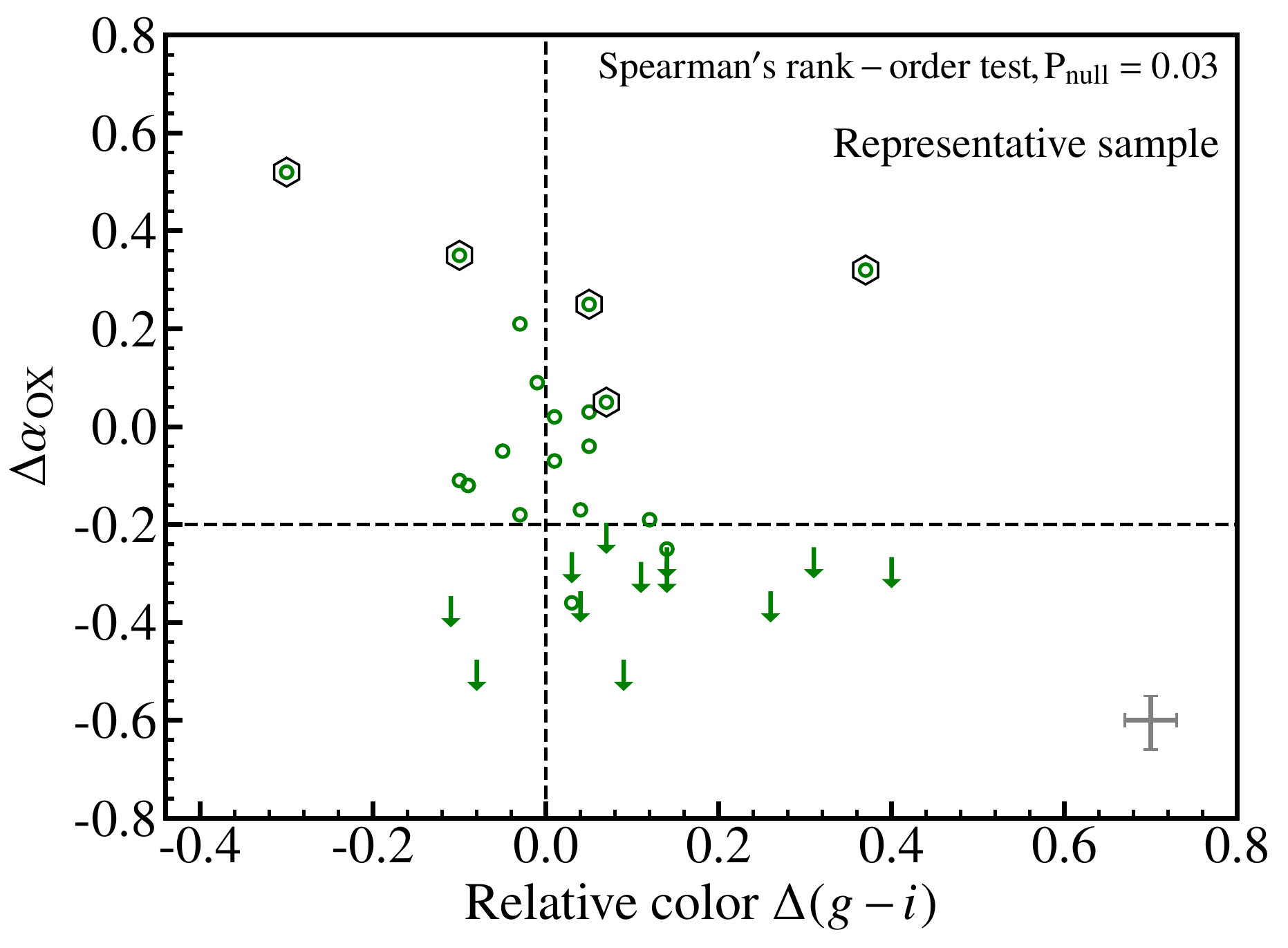}
\includegraphics[scale=0.46]{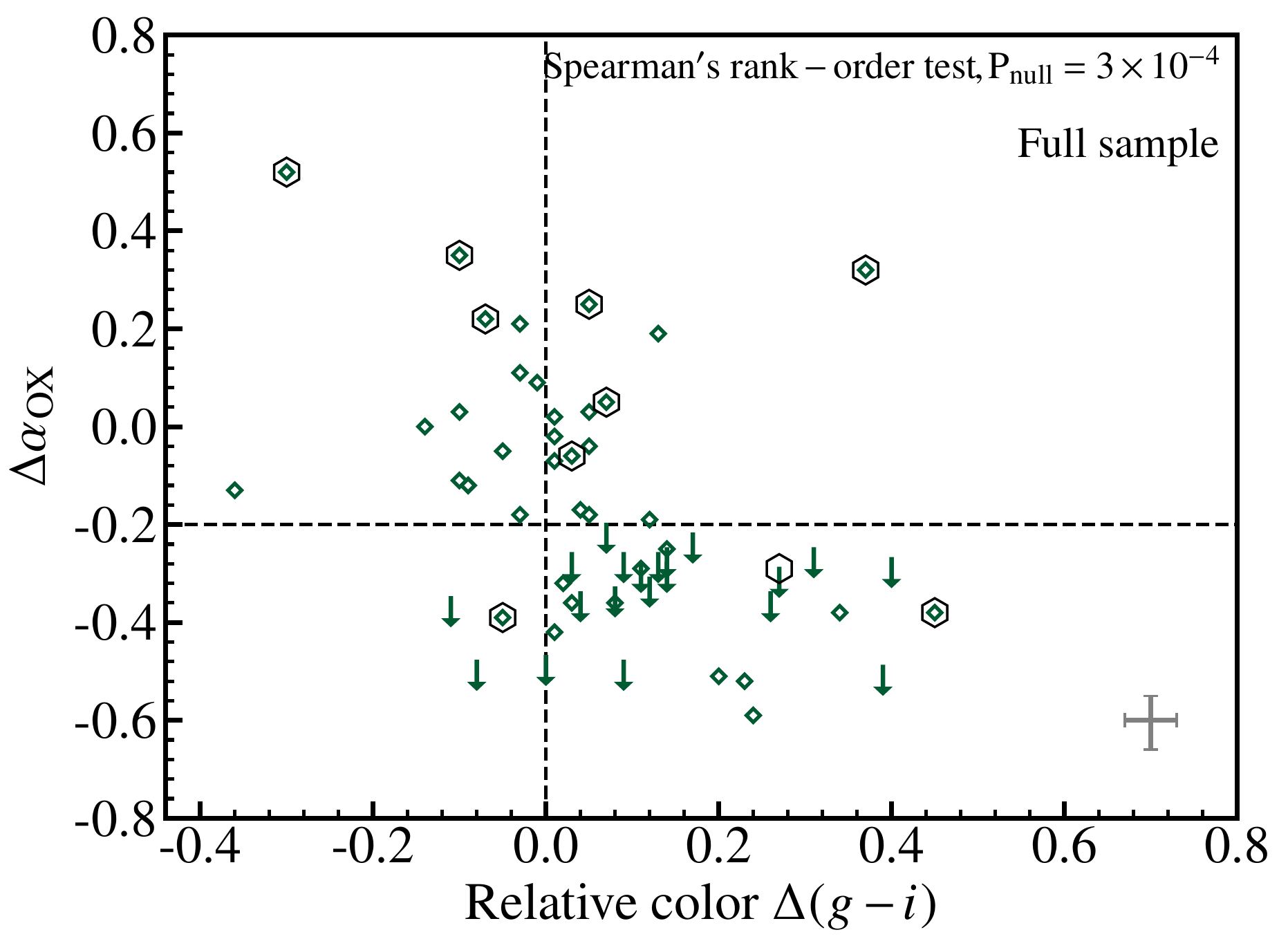}
}
\caption{$\Delta\alpha_{\rm OX}$ vs. relative color for the representative sample (left) and the full sample (right).
In cases of X-ray non-detections, the 90\% confidence upper limits of $\Delta\alpha_{\rm OX}$ are represented by the downward arrows.
The black hexagons encircling some of the WLQs indicate radio detections of these objects.
The median uncertainties of $\Delta\alpha_{\rm OX}$ and $\Delta(g-i)$ are shown as gray error bars in the bottom-right corners of both panels, and account only for the dominant error factors; i.e., the measurement errors of $\Delta\alpha_{\rm OX}$ and magnitudes.
The horizontal dashed lines display the 
division between \xray\ normal and \xray\ weak quasars adopted in this study.
The vertical dashed lines display the median color of SDSS quasars.
The test results from the Spearman rank-order test are listed for both panels, suggesting the correlation between $\Delta(g-i)$ and $\Delta\alpha_{\rm OX}$.
}
\label{gicorr}
\end{figure*}

\begin{figure*}
\centerline{
\includegraphics[scale=0.46]{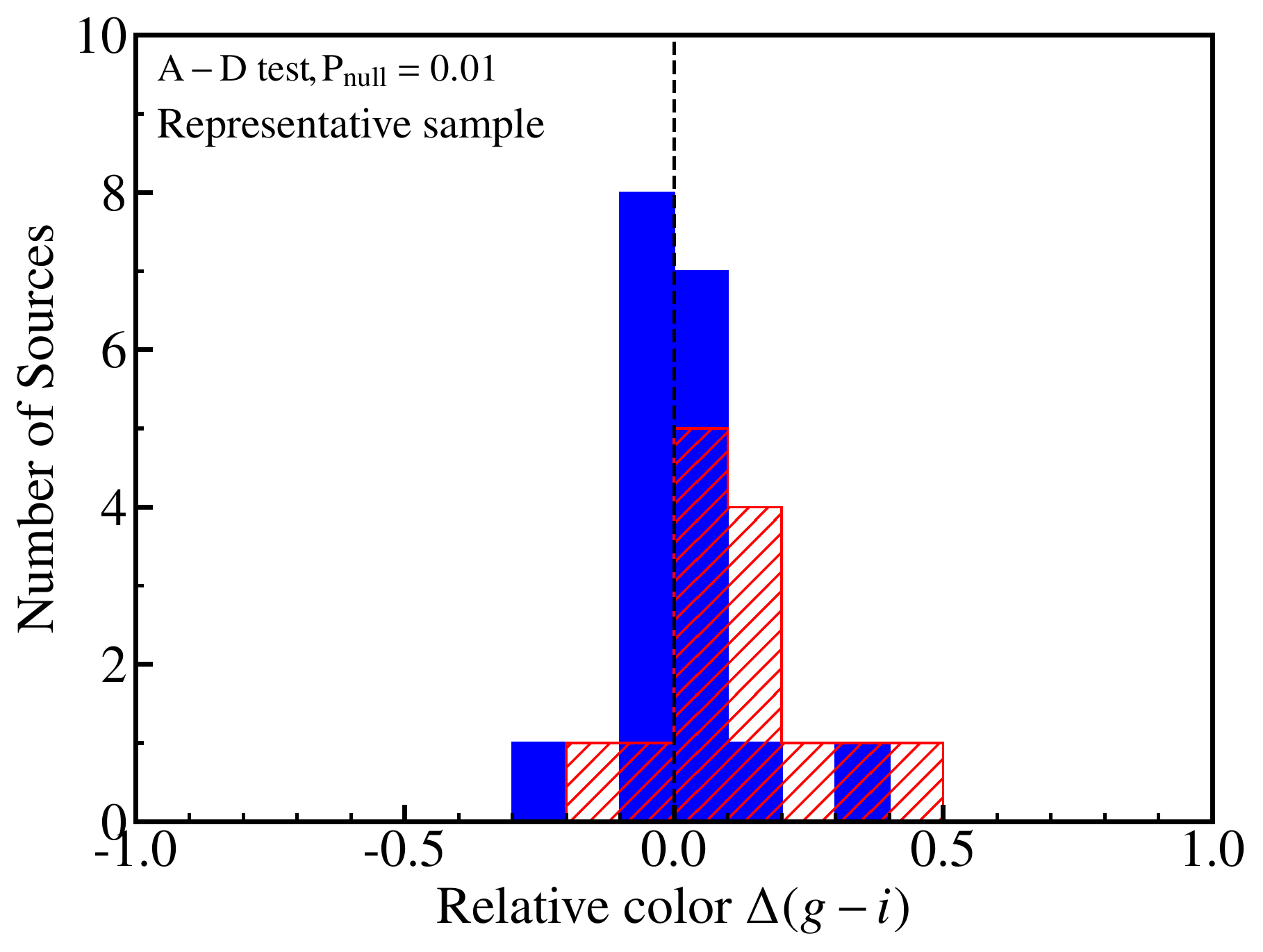}
\includegraphics[scale=0.46]{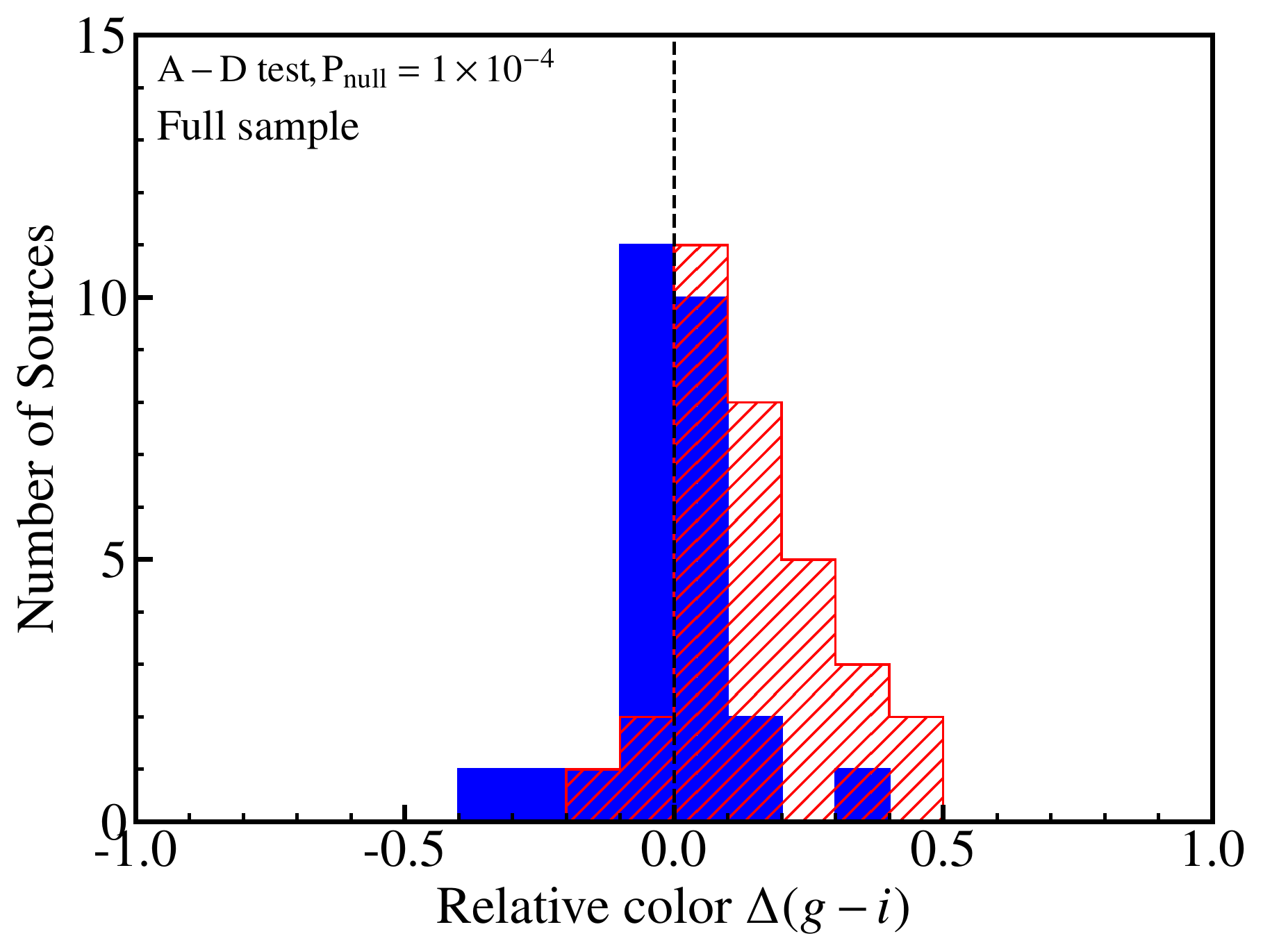}
}
\caption{Distributions of the relative color, $\Delta(g-i)$, for
the representative sample (left) and the full sample (right).
The hatched red histograms represent \xray\ weak quasars, while the solid blue histograms are the \xray\ normal quasars.
The vertical dashed line in each panel indicates a $\Delta(g-i)$ of zero.
The probabilities from the A-D test assessing whether the two distributions are drawn from the same population are displayed in both panels, suggesting the difference between the $\Delta(g-i)$ distributions of \xray\ weak WLQs and \xray\ normal WLQs for both the representative sample and the full sample.
In general, \xray\ weak WLQs are redder than \xray\ normal WLQs (and typical SDSS quasars).}
\label{gi}
\end{figure*}

While \citet{Gallagher2005} found no correlation between $\Delta(g-i)$ and $\Delta\alpha_{\rm OX}$ with a quasar sample of either large or small \iona{C}{iv} blueshifts, \citetalias{Luo2015} found the correlation to be significant among WLQs that typically have large \iona{C}{iv} blueshifts as well as small \iona{C}{iv} REWs.
As can be seen from Figure~\ref{gicorr}, the null probability that there is no correlation between $\Delta(g-i)$ and $\Delta\alpha_{\rm OX}$ is small.\footnote{The correlation becomes even more significant if the radio-detected objects are dropped. For the representative sample, the $P_{\rm null}$ of the Spearman rank-order test turns out to be 0.02; for the full sample, the $P_{\rm null}$ turns out to be $2 \times 10^{-4}$.} 
In the Peto-Prentice test table (Table~\ref{pptable}), the relative color $\Delta(g-i)$ is found to be the most-powerful UV tracer: \xray\ weak WLQs have redder $\Delta(g-i)$ than \xray\ normal WLQs at nearly the 4$\sigma$ significance level ($3.9\sigma$) in the full sample. Since there is no censoring for $\Delta(g-i)$, the Peto-Prentice test reduces to Gehan's Wilcoxon test \citep[e.g.][]{Lavalley1992}. 
In the representative sample, the median $\Delta(g-i)$ color among \xray\ weak WLQs is $0.10\pm 0.04$ (the 1$\sigma$ uncertainty is derived from bootstrap resampling), and it is $0.00 \pm 0.03$ among \xray\ normal WLQs. In the full sample, $\Delta(g-i)$ is $0.11\pm 0.02$ for \xray\ weak WLQs, and $0.00\pm 0.03$ for \xray\ normal WLQs.
The distributions of $\Delta(g-i)$ among \xray\ weak and \xray\ normal WLQs in both the representative sample and the full sample are presented in Figure~\ref{gi} with the results from the Anderson-Darling test (A-D test; \citealt{AD1952}; applicable when there is no censoring) in the SciPy package \citep{Scipy} listed, also showing the clear distinction.
Thus, the relative color $\Delta(g-i)$ appears to be an effective UV tracer of \xray\ weakness. 
As mentioned in Section~\ref{ssec-motivation}, this is plausibly consistent with the thick inner-disk model.

\subsubsection{Fe {\sc ii} REW}

\begin{figure*}
\centerline{
\includegraphics[scale=0.46]{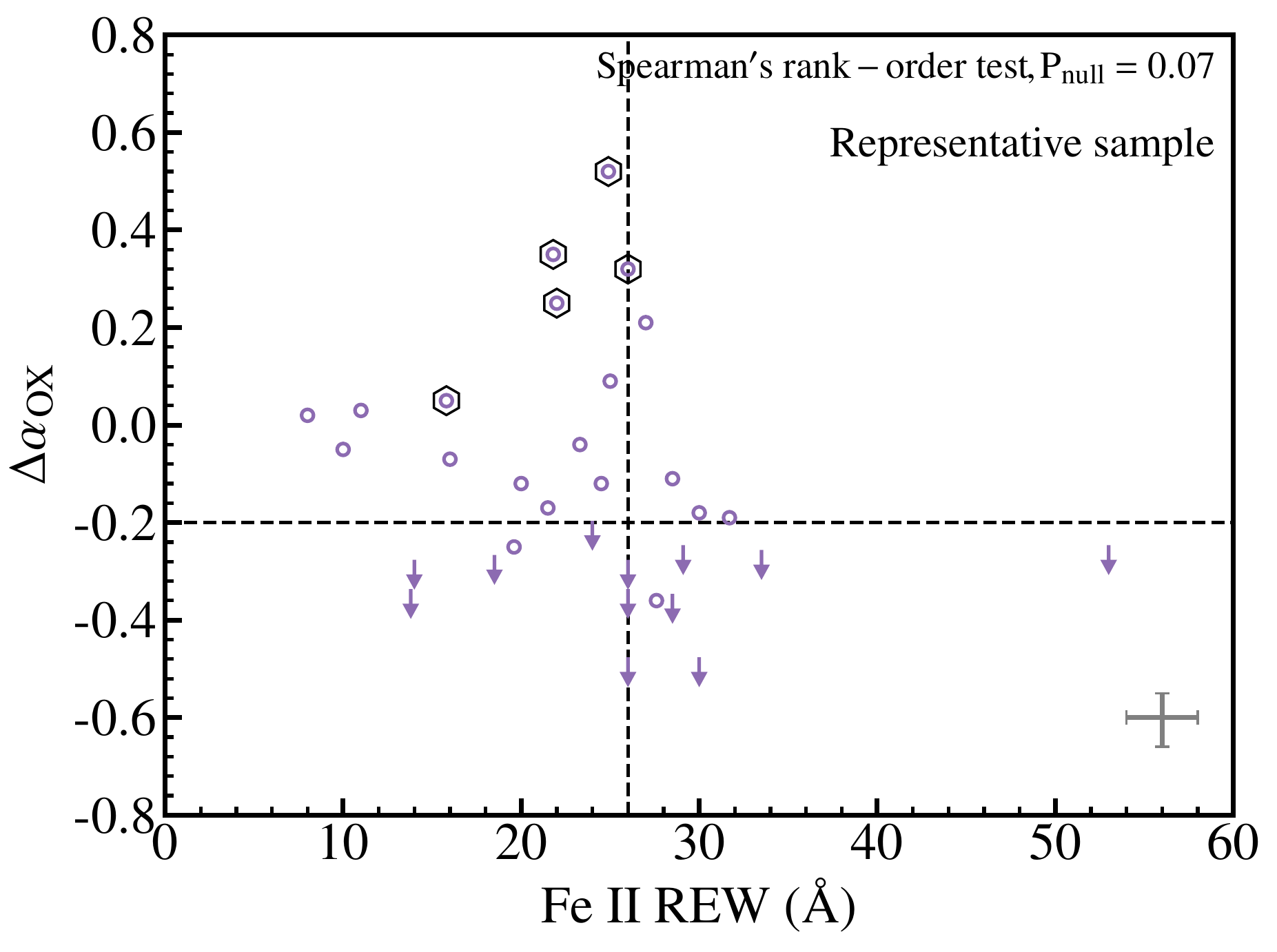}
\includegraphics[scale=0.46]{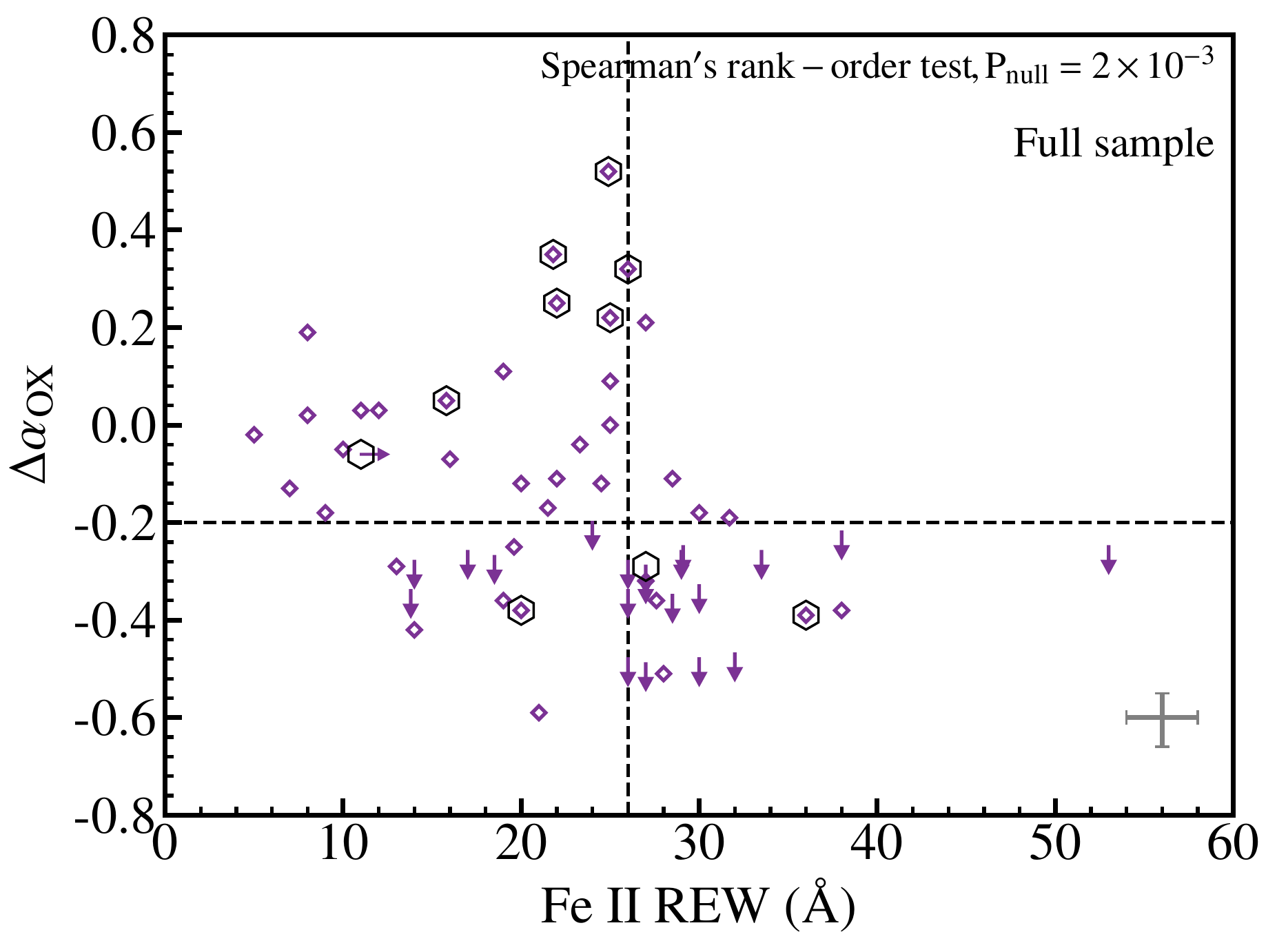}
}
\caption{$\Delta\alpha_{\rm OX}$ vs. \iona{Fe}{ii} REW for the representative sample (left) and the full sample (right).
In cases of X-ray non-detections, the 90\% confidence upper limits of $\Delta\alpha_{\rm OX}$ are represented by the downward arrows.
The black hexagons encircling some of the WLQs indicate radio detections of these objects.
The object in the full sample that only has limited spectral coverage of {Fe}{ \sc ii} (see the note of Table~3 in \citetalias{Luo2015}) is represented by the rightward arrow in the right panel.
The median uncertainties of $\Delta\alpha_{\rm OX}$ and {Fe}{ \sc ii} REW are shown as gray error bars in the bottom-right corners of both panels, and account only for the dominant error factors; i.e., the measurement errors.
The horizontal dashed lines display the 
division between \xray\ normal and \xray\ weak quasars adopted in this study.
The vertical dashed lines indicate the \iona{Fe}{ii} REW of the composite SDSS spectrum from \citet{Vandenberk2001}.
The test results from the Spearman rank-order test are listed for both panels, suggesting that while the correlation between \iona{Fe}{ii} REW and $\Delta\alpha_{\rm OX}$ is significant in the full sample, it is not significant in the representative sample.
}
\label{feiicorr}
\end{figure*}

\iona{Fe}{ii} REW was raised as another spectral diagnostic of \xray\ weak WLQs together with $\Delta(g-i)$ in \citetalias{Luo2015}.
The correlation between \iona{Fe}{ii} REW and $\Delta\alpha_{\rm OX}$ is significant in the full sample, but not significant in the representative sample (see Figure~\ref{feiicorr} with the Spearman rank-order test results listed).
One possibility for the significant correlation in the full sample is that it is artificial and simply driven by complex selection effects associated with the full sample (in \citetalias{Luo2015}, nearly half of the WLQs studied were originally selected and targeted with additional requirements for strong \iona{Fe}{ii} emission). Alternatively, the correlation found in the full sample may be real but simply not detectable in the representative sample owing to its smaller size. By randomly sampling 32 (the number of WLQs in the representative sample) objects from WLQs in the full sample, we estimated that the probability of getting significant Spearman rank-order test results ($P< 0.05$) given the current sample size is $\approx69.8\%$, derived from the empirical distribution of simulated test statistics. Thus, the small sample size could limit our ability of assessing \iona{Fe}{ii} REW as an effective UV tracer of \xray\ weakness among WLQs.

\begin{figure*}
\centerline{
\includegraphics[scale=0.48]{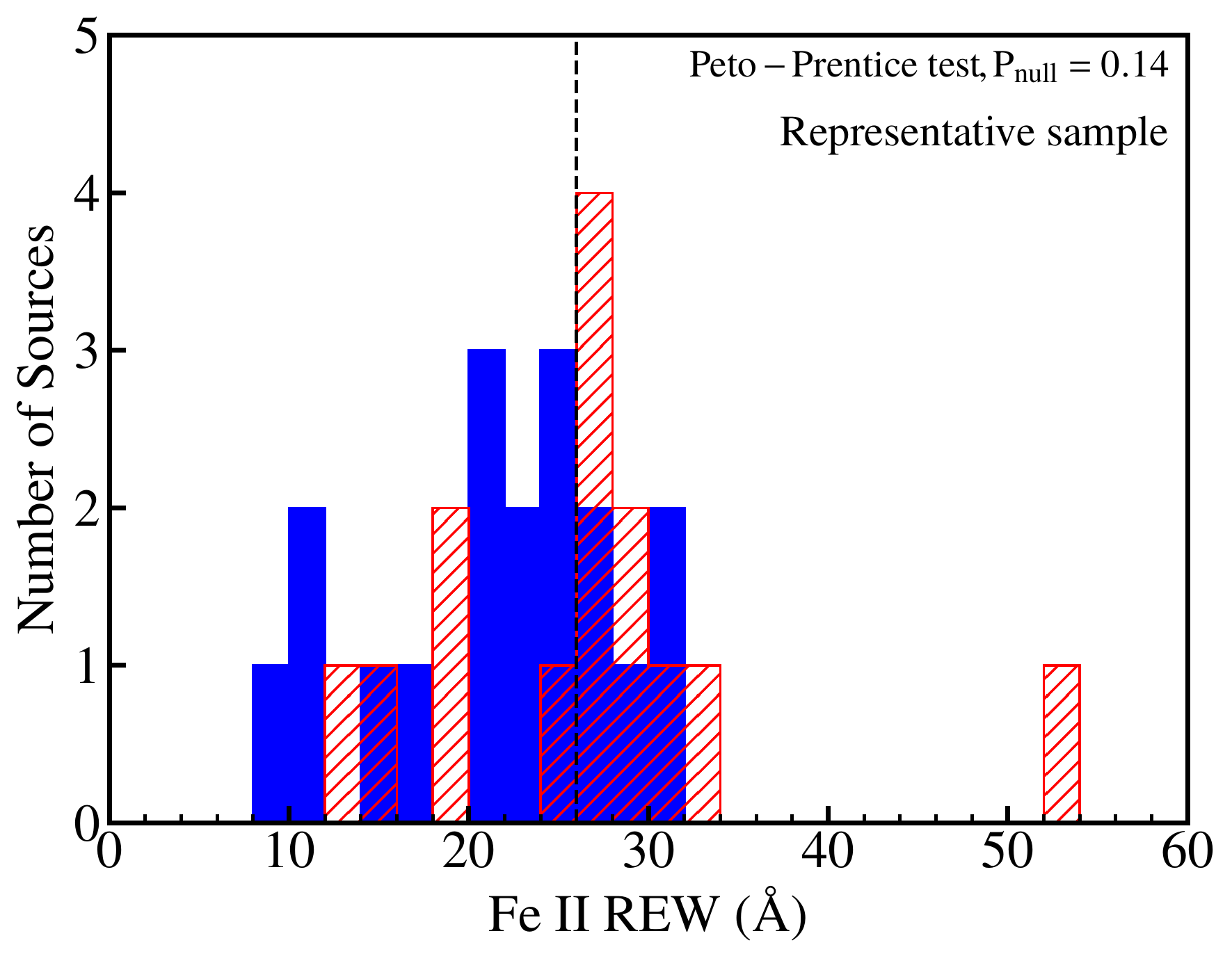}
\includegraphics[scale=0.48]{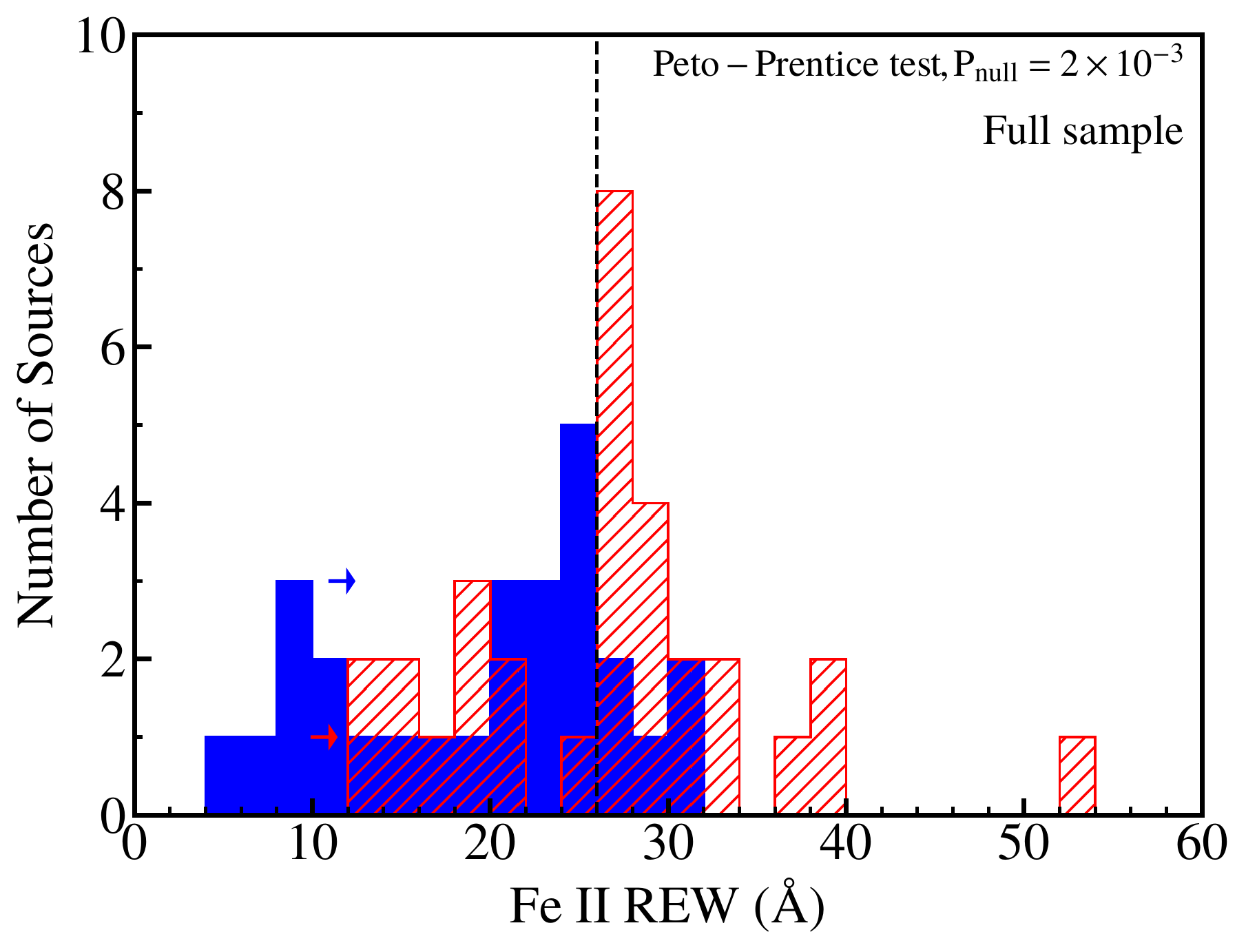}
}
\caption{Distributions of the \iona{Fe}{ii} REWs
for the representative sample (left) and the full sample (right).
The hatched red histograms represent \xray\ weak quasars, while the solid blue histograms are
\xray\ normal quasars.
The red/blue arrows represent limits of \iona{Fe}{ii} REWs measured from \xray\ weak/\xray\ normal quasars (see the note of Table~3, \citetalias{Luo2015}).
The vertical dashed lines indicate the \iona{Fe}{ii} REW of the composite SDSS spectrum from \citet{Vandenberk2001}.
The probabilities from the Peto-Prentice test are displayed in both panels, 
indicating that while the difference between the \iona{Fe}{ii} REW distributions of \xray\ weak WLQs and \xray\ normal WLQs is significant in the full sample, it is not significant in the representative sample.
}
\label{feii}
\end{figure*}

The results from the Peto-Prentice test (see Figure~\ref{feii} and Table~\ref{pptable}) also show that the difference in \iona{Fe}{ii} REW distributions between \xray\ normal WLQs and \xray\ weak WLQs is significant in the full sample but not significant in the representative sample. 
By randomly sampling 14 (the number of \xray\ weak WLQs in the representative sample) objects from all the \xray\ weak WLQs in the full sample and 18 (the number of \xray\ normal WLQs in the representative sample) objects from all the \xray\ normal WLQs in the full sample, we estimated that the probability of getting significant Peto-Prentice test results ($P< 0.05$) given the current sample size is $\approx61.1\%$. 
Clearly, only a larger unbiased WLQ sample with more detected \xray\ weak objects could help us investigate the potential of \iona{Fe}{ii} REW as a UV tracer of \xray\ weakness.

\subsubsection{Other emission features}
For the remaining spectral properties, including \iona{C}{iv} blueshift; \iona{C}{iv} FWHM; \iona{C}{iv} REW; 
and REWs of the \iona{Si}{iv}, $\lambda1900$, \iona{Fe}{iii}, and \iona{Mg}{ii} emission features, the Peto-Prentice test results in Table~\ref{pptable} reveal that the differences of property distributions between \xray\ weak WLQs and \xray\ normal WLQs can be found at the $0.08-0.9 \sigma$ level for the representative sample and the $0.03-1.4 \sigma$ level for the full sample. Thus, they are not effective UV tracers of \xray\ weakness. 
Leaving aside the ambiguous tracer \iona{Fe}{ii} REW, all the tested emission-line properties fail to distinguish significantly between \xray\ weak WLQs and \xray\ normal WLQs. 
In the context of Figure~\ref{schematic}, the lack of a connection between \iona{C}{iv} FWHM and \xray\ weakness is perhaps not surprising, since \citet{Runnoe2013} have shown that  \iona{C}{iv} FWHM is not a reliable orientation indicator. 

\subsubsection{Assessments over a broader range of C {\sc iv} REW and blueshift}
Although \iona{C}{iv} REW has worked effectively as the primary 
selector of our WLQs, it does not trace \xray\ weakness 
by itself particularly well among WLQs. For quasars with 
larger values of \iona{C}{iv} REW (primarily \hbox{15--100~\AA}), 
however, \citet{Gibson2008} show that $\Delta\alpha_{\rm OX}$ 
is correlated with \iona{C}{iv} REW at $>99.9$\% confidence. 
Furthermore, for such quasars, \citet{Gibson2008} find that 
$\Delta\alpha_{\rm OX}$ is correlated with the measured \iona{C}{iv} 
blueward wavelength shift at \hbox{97--99\%} confidence. It 
is therefore of interest to consider connections between 
$\Delta\alpha_{\rm OX}$ and \iona{C}{iv} properties more broadly
when our WLQ sample is combined with the \citet{Gibson2008} sample. 

\begin{figure*}
\centerline{
\includegraphics[scale=0.48]{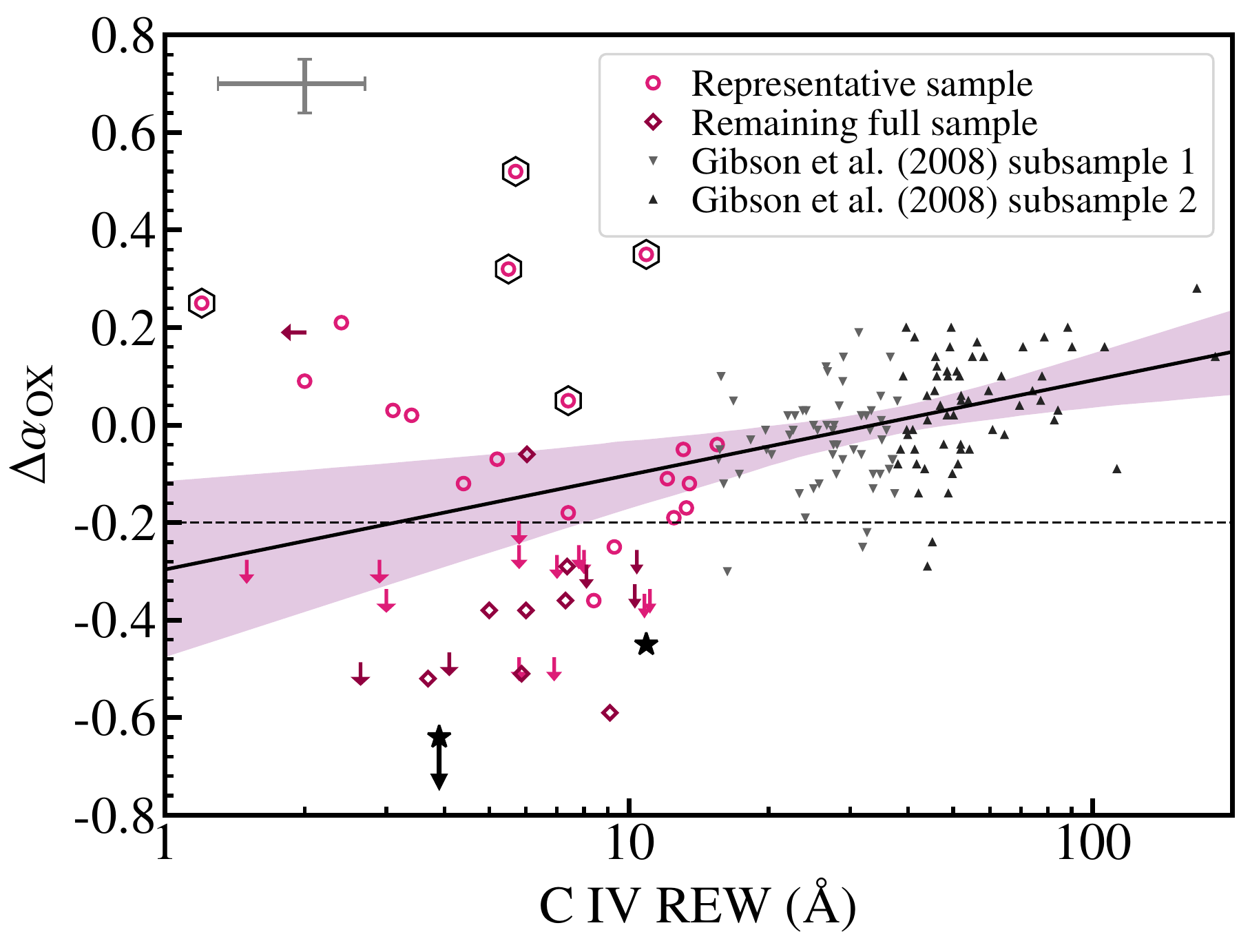}
\includegraphics[scale=0.48]{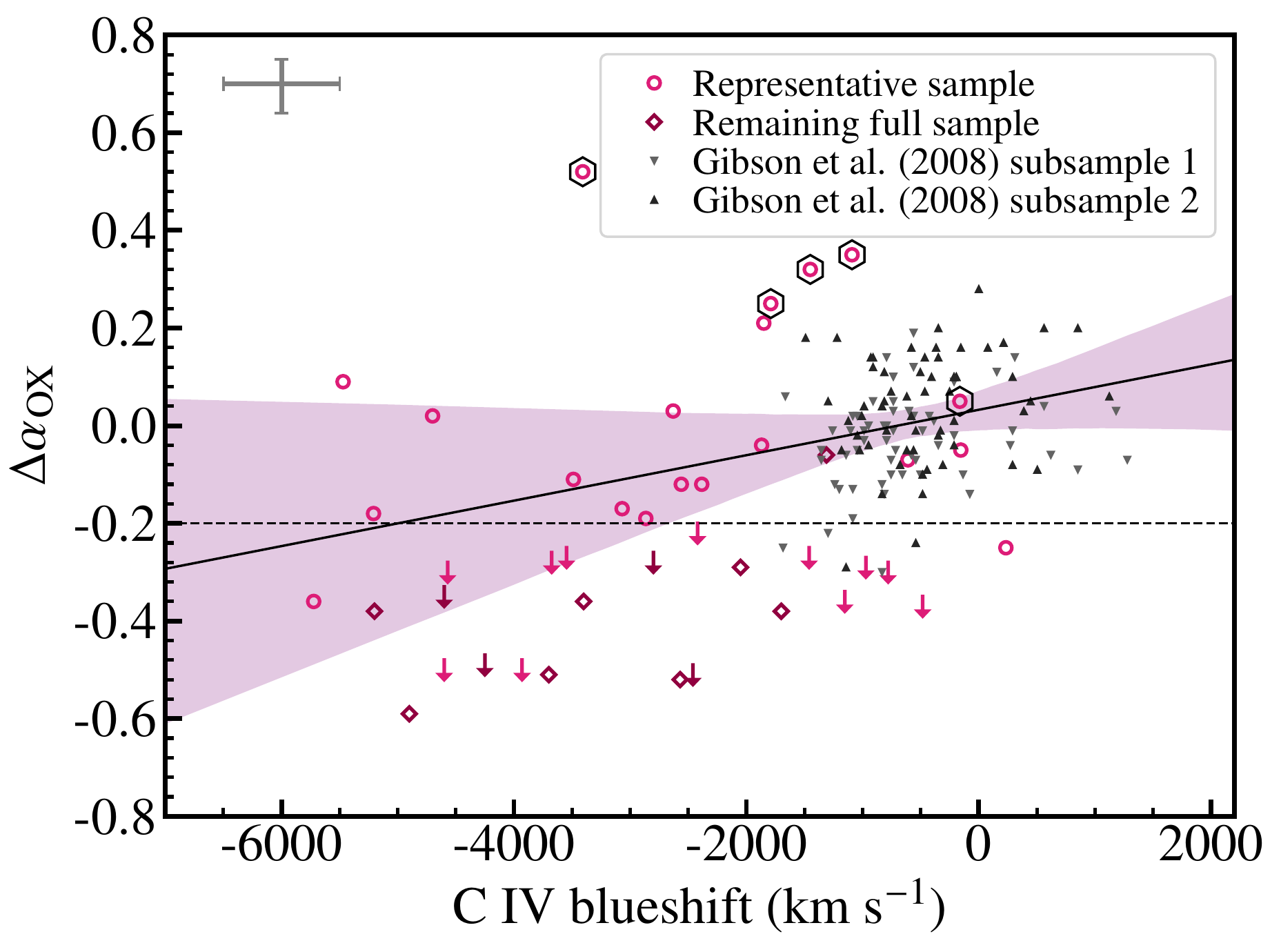}
}
\caption{$\Delta\alpha_{\rm OX}$ vs. \iona{C}{iv} REW (left) and $\Delta\alpha_{\rm OX}$ vs. \iona{C}{iv} blueshift (right).
The representative sample, the remaining part of the full sample (which consists of WLQs that are not included in the representative sample), and two subsamples of the \citet{Gibson2008} Sample B quasars are distinguished using symbol types as labeled. 
In cases of X-ray non-detections, the 90\% confidence upper limits of $\Delta\alpha_{\rm OX}$ are represented by the downward arrows.
The leftward arrow in the left panel presents the $3 \sigma$ upper limit of {C}{ \sc iv} REW.
The black hexagons encircling some of the WLQs indicate radio detections of these objects.
In the left/right panel, the median measurement errors of $\Delta\alpha_{\rm OX}$ and {C}{ \sc iv} REW/{C}{ \sc iv} blueshift for WLQs are shown as gray error bars in the upper-left corner.
The two black stars in the $\Delta\alpha_{\rm OX}$ vs. \iona{C}{iv} REW (left) panel represent the \xray\ stacking results: the one with an arrow represents the 90\% confidence upper limit of $\Delta\alpha_{\rm OX}$ derived from the \xray\ weak WLQs in the extreme subsample, and the one without an arrow represents the stacked mean $\Delta\alpha_{\rm OX}$ of \xray\ weak WLQs in the bridge subsample, each plotted at the median \iona{C}{iv} REW of WLQs in the subsample.
In both panels, the black solid line represents the posterior median estimate of the regression line obtained though fitting our ``ultra-clean'' version of the \citet{Gibson2008} Sample B quasars following the method in \citet{Kelly2007}, and the shaded region represents the $3\sigma$ pointwise confidence intervals on the regression line.
The horizontal dashed lines indicate the division between \xray\ normal and \xray\ weak quasars adopted in this study.}
\label{civgibson}
\end{figure*}

Figure~\ref{civgibson} shows $\Delta\alpha_{\rm OX}$ vs.\  
\iona{C}{iv} REW and \iona{C}{iv} blueshift. The relation 
between $\Delta\alpha_{\rm OX}$ and \iona{C}{iv} REW is 
apparent for \iona{C}{iv} REWs of \hbox{15--100~\AA}, although
even in this range the relation has considerable scatter
and thus the slope of the relation is poorly constrained. 
At lower values of \iona{C}{iv} REW, below $\approx 15$~\AA, 
the scatter of $\Delta\alpha_{\rm OX}$ abruptly becomes very large, 
so that one can no longer identify a clear relation between 
$\Delta\alpha_{\rm OX}$ and \iona{C}{iv} REW (and, indeed, 
the Spearman rank-order test finds no significant correlation
in this regime). Some WLQs do not appear to be \xray\ weak 
and may even be \xray\ stronger than expected from an extrapolation 
of the $\Delta\alpha_{\rm OX}$ vs. \iona{C}{iv} REW relation 
to small values of \iona{C}{iv} REW. On the other hand, 
stacking analyses show that many of the WLQs lie far below the 
extrapolation of the $\Delta\alpha_{\rm OX}$ vs.\ \iona{C}{iv} 
REW relation (see the black stars in Figure~\ref{civgibson}).
In the context of the shielding model shown in Figure~\ref{schematic}, 
this $\Delta\alpha_{\rm OX}$ diversity, corresponding to a 
range of $>100$ in relative \xray\ vs. optical/UV flux, is
largely attributable to high-inclination objects being obscured 
by the thick inner disk while low-inclination objects are not. 
However, the fact that some WLQs appear \xray\ strong relative 
to the extrapolated $\Delta\alpha_{\rm OX}$ vs.\ \iona{C}{iv} 
REW relation is puzzling. Several factors may be relevant 
to explaining this result. First, as noted above, the relation 
established at \iona{C}{iv} REW of \hbox{15--100~\AA} has
significant uncertainty in its slope, and thus its 
extrapolation below 15~\AA\ is uncertain on basic statistical 
grounds. Furthermore, it is not entirely clear that the 
relation can be appropriately extrapolated outside the domain 
where it was derived by a factor of \hbox{2--15} times 
in \iona{C}{iv} REW. Finally, we note that most of the 
strongest outliers toward positive $\Delta\alpha_{\rm OX}$ 
are radio-detected objects that may have some jet-linked 
contribution to their \xray\ fluxes (see Section~\ref{sec-radio} for 
further discussion). A large-scale study of the 
$\Delta\alpha_{\rm OX}$ vs.\ \iona{C}{iv} REW relation 
utilizing archival data would be valuable for clarifying
these issues. 

The plot of $\Delta\alpha_{\rm OX}$ vs.\ \iona{C}{iv} 
blueshift in Figure~\ref{civgibson} does not show any
tight correlation over blueshifts ranging from 
$-6000$~km~s$^{-1}$ to $+1000$~km~s$^{-1}$. The scatter 
in $\Delta\alpha_{\rm OX}$ at a given \iona{C}{iv} 
blueshift for WLQs is clearly large. The 
uncertainty of the slope of the $\Delta\alpha_{\rm OX}$ 
vs.\ \iona{C}{iv} blueshift relation for the 
\citet{Gibson2008} quasars is large, making
extrapolation to the large blueshifts of WLQs highly
uncertain. 

\section{Summary and Future Work}
\subsection{Summary of main results}
We have presented \xray\ and multiwavelength analyses of 32 WLQs that have been selected in an unbiased and consistent manner, thereby bridging the observational gap between WLQs and typical quasars and connecting their \xray\ properties.
The main points from this paper are the following:
\begin{enumerate}
\item
We selected 20 representative WLQs with \hbox{\iona{C}{iv} REW $\approx 5-15$~\AA} and obtained \chandra~observations of 14 of these having no sensitive archival \xray\ observations. \xray\ photometric analyses and UV emission-line measurements were performed for the 20 objects.
After combining this dataset with another 12 representative WLQs that have \iona{C}{iv} REW $<5$~\AA, we re-divided these 32 WLQs into two subsamples (extreme/bridge) based on their \iona{C}{iv} REW with equal numbers of objects (16/16 with \iona{C}{iv} REW below/above 7 \AA) to probe the relationships of UV and \xray\ properties.
See Sections \ref{sec-sp}, \ref{sec-Xray}, and \ref{sec-uvm}.
\item
We calculated $\Delta\alpha_{\rm OX}$ for our \chandra\ Cycle 17 sample objects and the archival objects that had not been previously analyzed. We also performed \xray\ stacking analyses to derive the average properties of our \xray\ weak WLQs since their limited numbers of detected counts prevented us from analyzing them individually. All the WLQs in the representative sample have IR-to-UV SEDs similar to those of typical quasars, indicating that $\Delta\alpha_{\rm OX}$ can be appropriately used to assess the level of line-of-sight \xray\ emission relative to the longer wavelength SED. The representative WLQs are also radio-quiet ($R<10$), minimizing the influence from jets on the observed \xray\ properties.
See Sections \ref{sec-aox}, \ref{ssec-stack}, \ref{ssec-sed}, and \ref{sec-radio}.
\item
The hard power-law (0.5--8 keV, corresponding to \hbox{$\approx 2$--23~keV} in the rest frame) effective photon index, $\Gamma_{\rm eff}=1.19_{-0.45}^{+0.56}$, measured from \xray\ stacking of 14 \xray\ weak WLQs in the representative sample, suggests that the observed \xray\ weakness may be generally explained by heavy absorption due to small-scale shielding. This is consistent with the thick inner-disk model. See Section \ref{sec-nature}.
\item
A total of 7 out of 16 ($43.8^{+12.4}_{-11.0}$\%) WLQs in the bridge subsample are \xray\ weak, and this also applies to the extreme subsample. 
Comparing with typical quasars in \citet{Gibson2008} Sample~B, we established and quantified an overall trend, that is especially pronounced in the \hbox{10--20 \AA} range, demonstrating that quasars with weaker \iona{C}{iv} REW are more likely to be \hbox{X-ray} weak. This trend is strong; the fraction of \xray\ weak quasars varies by a factor of $\approx 13$ for \iona{C}{iv} REW ranging from \hbox{4--50~\AA}. The trend is expected from the shielding model, and thus these results generally support the model. As the inner accretion disk becomes thicker, ionizing radiation is increasingly blocked from reaching the \iona{C}{iv} BELR and, for random orientations, our line of sight to the central \xray\ emitting region is more likely to be blocked.
See Section \ref{sec-frac}.
\item
Among the UV continuum and emission-line properties examined that might be tracers of \xray\ weakness among WLQs, the relative SDSS color $\Delta(g-i)$ proved to be the most effective tracer. 
\xray\ weak WLQs have redder $\Delta(g-i)$ (at $3.9\sigma$ significance). 
See Section \ref{sec-UV}.
\end{enumerate}

\subsection{Future work}
There are a number of ways the results in this paper might
be extended. For example, the current sizes of our extreme
and bridge WLQ subsamples remain small, and these should be
enlarged via additional \hbox{X-ray} observations for improved
power in statistical hypothesis testing and trend quantification.
Such improved samples (selected with a lower $m_i$ threshold) would allow better quantification
of how the fraction of \hbox{X-ray} weak quasars depends upon
\iona{C}{iv} REW with implications for the thick inner-disk
model; see Figure \ref{fraction}(a). 
Note that the observed dependence is particularly strong for \iona{C}{iv} REW = 10--20 \AA, thus, increasing the sample size in this \iona{C}{iv} REW range may be particularly useful.
In addition to increasing the sizes of the
extreme and bridge subsamples, improving the data quality for
their constituent objects is also important. Presently,
many of the objects in these samples are \hbox{X-ray} undetected,
so that only limits on their $\Delta\alpha_{\rm OX}$ values
are available (and these $\Delta\alpha_{\rm OX}$ limits are often relatively loose).
While we have been able to recover useful
information from these undetected objects via survival
analysis and stacking, reducing the fraction of \hbox{X-ray}
upper limits would provide a clearer picture of the overall
population. For example, we would be able to examine the
distribution of $\Delta\alpha_{\rm OX}$ values, assessing
potential diversity within the WLQ population (e.g. the possible
presence of some intrinsically X-ray weak objects), and
our \hbox{X-ray} spectral analyses would also benefit from
the additional detected counts. We would furthermore be able to
investigate the general UV spectral tracers of \hbox{X-ray}
weakness more effectively via correlation analyses. 
Considering our stacking results, we should be able to detect most of the currently \hbox{X-ray}
undetected sources with exposure
times $\approx 5-10$ times longer than the present ones.
While such exposures would not be inexpensive, they
are not impracticable. 

Furthermore, in the construction of our WLQ samples, we deliberately
excluded quasars with significant \iona{C}{iv} absorption (e.g. BALs
and mini-BALs) and also those with very red UV spectra (see Section~\ref{ssec-select}).
While this was necessary and appropriate for this project, it would
now be useful to target such objects with \chandra\ and \xmm\
so that their \xray\ properties could be connected with those of
the WLQs in our current samples. This seems especially relevant
given that we find a connection between \xray\ weakness and red
UV color (Section~\ref{sec-UV}) and given that the fraction of quasars with
BALs likely rises with Eddington ratio (e.g. \citealt{Boroson2002,Ganguly2007}).
Ideally, one would like to study the fraction
of \xray\ weak quasars vs. \iona{C}{iv} REW relation (Figure \ref{fraction}(a)) with such
objects included, though this will likely be technically challenging
(e.g. measurements of the \iona{C}{iv} REW may be affected by \iona{C}{iv}
absorption and complex intrinsic reddening).

In addition, several of the main results in this
work have been derived by comparing our WLQ subsamples with 
the archival \hbox{X-ray} quasar sample of \citet{Gibson2008}
(specifically, their Sample~B). In the decade since Sample~B
was created, both the \hbox{X-ray} archives and underlying SDSS
quasar database have grown substantially. Further data mining
based on the current \hbox{X-ray} archives and SDSS database could
define a substantially larger version of this sample for more
statistically powerful comparisons with WLQs.

Finally, we note that our thick inner-disk model for WLQs has similarities with
models of other rapidly accreting systems, including stellar tidal disruption events
(TDEs; e.g. \citealt{Dai2018}) and ultraluminous X-ray sources (ULXs; e.g. \citealt{Kaaret2017}).
For example, \citet{Dai2018} suggest that the X-ray vs.\ optical fluxes of TDEs will have an
inclination-angle dependence that is qualitatively similar to what we previously
proposed for WLQs, due to obscuration of small-scale \xray\ emission by a thick
accretion disk and its associated outflow. Along these lines, one might
speculate that WLQs have a (slow) transient nature as well. Their SMBHs are
likely too massive typically to allow tidal disruptions of stars (e.g. see
Section~5.3 of \citetalias{Luo2015} for SMBH mass estimates), but captures of
$10^3$--$10^6$~M$_\odot$ nuclear molecular clouds are plausible and could
feed the SMBH at a super-Eddington rate for decades to millennia.
Generally, comparisons of the properties of WLQs, TDEs, and ULXs should be
productive for interpreting the nature of all three classes and suggesting
valuable new observations.


\section*{Acknowledgements}

We thank Fabio Vito and Shifu Zhu for helpful discussions.
QN and WNB acknowledge support from Chandra X-ray Center grant
GO6-17083X, the NASA ADP Program, and the Penn State ACIS
Instrument Team Contract SV4-74018 (issued by the Chandra
X-ray Center, which is operated by the Smithsonian Astrophysical
Observatory for and on behalf of NASA under contract NAS8-03060).
The Guaranteed Time Observations (GTO) for some of the quasars
studied were selected by the ACIS Instrument Principal
Investigator, Gordon P. Garmire, currently of the Huntingdon
Institute for X-ray Astronomy, LLC, which is under contract to
the Smithsonian Astrophysical Observatory; Contract SV2-82024.
BL acknowledges financial support from the National Key R\&D Program of China grant 2016YFA0400702 and the National Natural Science Foundation of China grant
11673010. PBH acknowledges support from NSERC, funding reference number 2017-05983. YS acknowledges support from an Alfred P. Sloan Research Fellowship and NSF grant AST-1715579. RMP acknowledges support from Curtin University through the Peter Curran Memorial Fellowship.


\bibliographystyle{mnras}
\bibliography{wlqbridgelib}



\clearpage
\begin{deluxetable}{lccccccccc}
\setlength{\tabcolsep}{1pt}
\tabletypesize{\footnotesize}
\tablewidth{0pt}
 
 \tablecaption{Summary of Utilized Quasar Samples}
 \tablehead{
 \colhead{Sample}  & \colhead{Number of}   & \colhead{Range of}   &\colhead{Median}  &\colhead{Range of}  & \colhead{Median}  & \colhead{Range of}  & \colhead{Median}  &
 \colhead{Relevant} &\colhead{Notes}  \\
 \colhead{Name} & \colhead{Sources}  & \colhead{\iona{C}{iv} REW}  & \colhead{\iona{C}{iv} REW}  & \colhead{Redshift}  & \colhead{Redshift}  & \colhead{$\log L_{\rm 2500~\textup{\AA}}$}  & \colhead{$\log L_{\rm 2500~\textup{\AA}}$}   & \colhead{Section}  & \colhead{}     \\
 \colhead{}  & \colhead{}   & \colhead{(\AA)} & \colhead{(\AA)}   & \colhead{}  & \colhead{}  & \colhead{(\mlum)} & \colhead{(\mlum)}  & \colhead{}  & \colhead{} \\
 \colhead{(1)}                   &
 \colhead{(2)}                   &
 \colhead{(3)}               &
 \colhead{(4)}               &
 \colhead{(5)}                  &
 \colhead{(6)}               &
 \colhead{(7)}               &
 \colhead{(8)}               &
 \colhead{(9)}                  &
 \colhead{(10)}               
}
 \startdata
 \vspace{0.5mm}
 Extreme Subsample\tablenotemark{a} & 16 & 0--7.0      & $3.9 \pm 0.9$ &  1.5--2.5 & $1.83 \pm 0.08$ & 30.95--31.75 & $31.28\pm 0.06$ & 2.2 & WLQ  \\
 \vspace{0.5mm}
 Bridge Subsample & 16 & 7.0--15.5    & $10.9 \pm 1.2$& 1.7--2.2 & $1.92 \pm 0.06$ & 31.15--31.89 & $31.40\pm 0.06$ &  2.2 & WLQ \\ 
 \vspace{0.5mm}
 Representative Sample\tablenotemark{b} & 32 & 0--15.5      & $7.0 \pm 0.9$  & 1.5--2.5 & $1.87 \pm 0.06$ & 30.95--31.89 & $31.35\pm 0.04$& 2.2& WLQ\\ 
 \vspace{0.5mm}
 Full Sample\tablenotemark{c}  & 63 & 0--15.5  & $6.5 \pm 0.7$  &  0.4--3.0& $1.80 \pm 0.05$ & 29.59--32.39 & $31.26\pm 0.05$& 2.3 & WLQ \\
 \vspace{0.5mm}
\citet{Gibson2008}  & 59 & 15.5--38.0     & $27.7 \pm 1.1$& 1.7--2.7 & $2.03\pm 0.06$ & 30.33-31.67 & $31.00 \pm 0.06$ & 2.2 & Typical quasars\\ 
 Subsample 1  &&&&&&&&&\\
 \vspace{0.5mm}
\citet{Gibson2008}  & 60 & 38.0--183.6& $50.5 \pm 1.4$& 1.7--2.7 & $1.89\pm 0.03$ & 30.39--31.46 &$30.86 \pm 0.05$ & 2.2 & Typical quasars\\
 Subsample 2 &&&&&&&&&
 \enddata
 \tablecomments{Cols. (1) and (2): Sample names utilized in this paper and the number of sources within the sample.
Cols. (3) and (4): Range of \iona{C}{iv} REW for each sample and median value with the bootstrapped error.
Cols. (5) and (6): Range of redshift for each sample and median value with the bootstrapped error.
Cols. (7) and (8): Range of logarithm of the monochromatic luminosity at rest-frame 2500~\AA\ for each sample and median value with the bootstrapped error.
Col. (9): Relevant sections describing the selection and constitution of each sample.
Col. (10): Comments for each sample.\newline
$\rm ^a$Most of the WLQs in the extreme subsample have been well-studied previously and thus are not included in our tables.
Detailed multiwavelength analyses of J082508.75+115536.3, J082722.73+032755.9, J084424.24+124546.5, J100517.54+331202.8, J115637.02+184856.5, J132809.59+545452.7, J134601.28+585820.2, J140710.26+241853.6, J141141.96+140233.9, J141730.92+073320.7, J153913.47+395423.4, and J172858.16+603512.7 can be found in \citetalias{Luo2015}; detailed multiwavelength analyses of J094533.98+100950.1 and J161245.68+511816.9 can be found in \citetalias{Wu2012}; detailed multiwavelength analyses of J090312.22+070832.4 can be found in \citetalias{Wu2011}. Only one object, J113949.39+460012.9 (which has an archival X-ray observation), has not been analyzed before, and it is thus listed in our tables along with WLQs in the bridge subsample.
\newline
$\rm ^b$Representative sample is the sum of the extreme subsample and bridge subsample.\newline
$\rm ^c$Full sample includes representative sample and other WLQs that are not selected in as systematic a manner. For the full sample, cols (3) and (4) only apply to the 46 objects with SDSS spectral coverage of \iona{C}{iv}.
}
 \label{summarytable}
\end{deluxetable}

\clearpage
\begin{landscape}
\begin{deluxetable}{lccccccccccc}
\setlength{\tabcolsep}{4pt}
\tabletypesize{\footnotesize}
\tablewidth{0pt}
 \tablecaption{X-ray Observations and Photometric Properties of WLQs in the Bridge Subsample}
 \tablehead{
 \colhead{Object Name}                &
 \colhead{RA}                                &
 \colhead{Dec}                                &
 \colhead{Redshift}                        &
 \colhead{Observation}                  &
 \colhead{Observation}                  &
 \colhead{Exposure}                   &
 \colhead{Soft Band}                   &
 \colhead{Hard Band}                   &
 \colhead{Band}                   &
 \colhead{$\Gamma_{\rm eff}$}                   &
 \colhead{Comment}                   \\
 \colhead{(J2000)}   &
 \colhead{(deg)}   &
 \colhead{(deg)}   &
 \colhead{}   &
 \colhead{ID}   &
 \colhead{Start Date}   &
 \colhead{Time (ks)}   &
 \colhead{(0.5--2~keV)}   &
 \colhead{(2--8~keV)}   &
 \colhead{Ratio}                   &
 \colhead{}                   &
 \colhead{}   \\
 \colhead{(1)}         &
 \colhead{(2)}         &
 \colhead{(3)}         &
 \colhead{(4)}         &
 \colhead{(5)}         &
 \colhead{(6)}         &
 \colhead{(7)}         &
 \colhead{(8)}         &
 \colhead{(9)}         &
  \colhead{(10)}         &
 \colhead{(11)}         &
 \colhead{(12)}  \\
 \noalign{\smallskip}\hline\noalign{\smallskip}
 \multicolumn{12}{c}{Chandra Cycle 17 Objects} 
}
 \startdata
 $       080040.77+391700.4$& 120.169914 & 39.283451 &$ 1.777$&$  18110$&    2015 Dec 17&$ 3.1$&$  8.3_{-2.9}^{+4.1}$&$  4.4_{-2.1}^{+3.5}$&$ 0.52_{-0.31}^{+0.50}$&$  1.6_{-0.5}^{+0.7}$&     ... \\
 $       095023.19+024651.7$& 147.596664 & 2.781048 &$ 1.882$&$  18118$&    2016 Jan 28&$ 4.8$&$                      <4.0$&$                      <4.2$&$                              ...$&$                          ...$&     ... \\
 $       101209.62+324421.4$& 153.040115 & 32.739295 &$ 1.985$&$  18113$&    2016 Feb 11&$ 4.2$&$  7.3_{-2.7}^{+3.9}$&$  6.5_{-2.6}^{+4.0}$&$ 0.89_{-0.49}^{+0.73}$&$   1.1_{-0.4}^{+0.6}$&    ... \\
 $       101945.26+211911.0$& 154.938583 & 21.319723 &$ 1.842$&$  18109$&    2015 Dec 29&$ 3.1$&$11.5_{-3.4}^{+4.6}$&$  4.3_{-2.1}^{+3.5}$&$ 0.38_{-0.22}^{+0.34}$&$   1.8_{-0.5}^{+0.6}$&    ...\\
 $       110409.96+434507.0$& 166.041519 & 43.751972 &$ 1.804$&$  18119$&    2016 Jan 4&$    4.6$&$                     <2.4$&$                      <2.5$&$                              ...$&$                           ...$&     ...\\
 $       122048.52+044047.6$&185.202179 &  4.679912 &$ 1.737$&$  18112$&    2016 Jan 21&$ 3.4$&$  7.3_{-2.7}^{+3.9}$&$                      <4.3$&$                        <0.59$&$                       >1.4$&     ...\\
 $       122311.28+124153.9$& 185.797028 & 12.698329 &$ 2.068$&$  18115$&    2016 Feb 11&$ 4.5$&$                      <4.0$&$                      <4.2$&$                              ...$&$                           ...   $&                      ...\\
 $       122855.90+341436.9$& 187.232941 & 34.243595 &$ 2.147$&$  18111$&    2016 Feb 15&$ 3.9$&$                      <4.1$&$                      <2.5$&$                              ...$&$                           ...$&     ...\\
 $       124516.46+015641.1$& 191.318604 & 1.944762 &$ 2.006$&$  18114$&    2016 Feb 22&$ 4.3$&$  9.3_{-3.1}^{+4.3}$&$12.0_{-3.6}^{+4.9}$&$1.29_{-0.57}^{+0.79}$&$    0.9_{-0.4}^{+0.4}$&                      ...\\
 $       140701.59+190417.9$& 211.756653 & 19.071661 &$ 2.004$&$  18117$&    2016 Apr 5&$    5.0$&$  2.0_{-1.3}^{+2.8}$&$                       <2.5$&$                       <1.42$&$                      >0.8$&                      ...\\
 $       150921.68+030452.7$& 227.340347& 3.081327 &$ 1.808$&$  18107$&    2015 Dec 28&$ 3.1$&$18.7_{-4.4}^{+5.6}$&$12.1_{-3.6}^{+4.9}$&$ 0.64_{-0.24}^{+0.32}$&$   1.4_{-0.3}^{+0.4}$&                      ...\\
 $       155621.31+112433.2$& 239.088806 & 11.409227 &$ 1.730$&$  18108$&    2016 May 12&$ 3.1$&$ 7.3_{-2.7}^{+3.9}$&$   9.9_{-3.2}^{+4.5}$&$ 1.36_{-0.67}^{+0.97}$&$   0.9_{-0.4}^{+0.5}$&                      ...\\
 $       163810.07+115103.9$& 249.541992 & 11.851092 &$ 1.983$&$  18116$&    2016 Feb 17&$ 4.1$&$                       <2.4$&$                     <2.5$&$                              ...$&$                          ...  $&                      ...\\
 $       232519.33+011147.8$& 351.330566 & 1.196628 &$ 1.727$&$  18120$&    2016 Aug 15&$ 4.3$&$  7.3_{-2.7}^{+3.9}$&$   8.8_{-3.0}^{+4.4}$&$ 1.21_{-0.61}^{+0.89}$&$  1.0_{-0.4}^{+0.5}$&                      ...\\
 \noalign{\smallskip}\hline\noalign{\smallskip}
 \multicolumn{12}{c}{Archival data} \\
 \noalign{\smallskip}\hline\noalign{\smallskip}
 $       123326.03+451223.0$& 188.358459 & 45.206402 &$ 1.966$&$           26747$&    1990 Nov 27&$   0.5$&$      21.6_{-6.2}^{+6.2}$\tablenotemark{a}&$                          ...$&$                              ...  $&$         ...$&  \textit{RASS2RXS}\\
$        215954.45-002150.1$& 329.976929 & -0.363942 &$ 1.965$&$           11509$&     2009 Sep 20&$   7.9$&$106.3_{-10.5}^{+11.6}$&$ 39.5_{-6.6}^{+7.8}$&$ 0.37_{-0.07}^{+0.08}$&$  1.9_{-0.2}^{+0.1}$&  Cycle~11\\
$       113949.39+460012.9^* $& 174.955795 & 46.003604 &$ 1.859$&$ 9155100004$&   2008 May 29&$0.008$&$        4.0_{-2.0}^{+2.0}$\tablenotemark{b}&$                          ...$&$                              ...$&$       ...$&   \textit{XMMSL2}
 \enddata
 \tablecomments{
Col. (1): Object name in the J2000 coordinate format. The object marked with ``$^*$'' comes from the extreme subsample.
Cols. (2)--(3): The SDSS position in decimal degrees.
Col. (4): Redshift adopted from \citet{Hewett2010}.
Cols. (5)--(6): The ID and start date of \chandra\ observations.
Col. (7): Effective exposure time in the full band (0.5--8~keV) with background flares cleaned.
Cols. (8)--(9): Source counts (aperture-corrected) in the soft band (0.5--2~keV) and hard band (2--8~keV). 
If the source is undetected in this band, an upper limit of counts at a 90\% confidence level is listed.
Col. (10): Ratio of the hard-band and soft-band counts.  ``...'' indicates that the source is not detected in both bands.
Col. (11): Effective power-law photon index in the \hbox{0.5--8~keV} band. ``...'' indicates that $\Gamma_{\rm eff}$ cannot be constrained.
Col. (12): Comments for the object.\newline
$\rm ^a$For \textit{RASS2RXS} (Second ROSAT All-Sky Survey Point Source Catalog) data, the soft band is 0.1--2.4~keV.\newline
$\rm ^b$For \textit{XMMSL2} (Second \xmm~Slew Survey Catalog) data, the soft band is 0.2--2~keV.
}
 \label{xraytable}
\end{deluxetable}
\end{landscape}

\clearpage
\begin{landscape}
 \begin{deluxetable}{lccccccccccccc}
  \tabletypesize{\footnotesize}
  \tablewidth{0pt}
  \tablecaption{X-ray and Optical Properties of Quasars in the Bridge Subsample}
  \tablehead{
\colhead{Object Name}                   &
\colhead{$M_{i}$}                   &
\colhead{$N_{\rm H,Gal}$}                   &
\colhead{Count Rate}                   &
\colhead{$F_{\rm X}$}                   &
\colhead{$f_{\rm 2~keV}$}                   &
\colhead{$\log L_{\rm X}$}                   &
\colhead{$f_{\rm 2500~\textup{\AA}}$}                   &
\colhead{$\log L_{\rm 2500~\textup{\AA}}$}                   &
\colhead{$\alpha_{\rm OX}$}                   &
\colhead{$\Delta\alpha_{\rm OX}(\sigma)$}                   &
\colhead{$f_{\rm weak}$}                  &
\colhead{$\Delta(g-i)$}  &
\colhead{$R$}  \\
\colhead{(J2000)}   &
\colhead{}   &
\colhead{} & 
\colhead{(0.5--2~keV)}   &
\colhead{(0.5--2~keV)}   &
\colhead{}   &
\colhead{(2--10 keV)}   &
\colhead{}   &
\colhead{}   &
\colhead{}   &
\colhead{}   &
\colhead{}   &
\colhead{}   &
\colhead{}   \\
\colhead{(1)}         &
\colhead{(2)}         &
\colhead{(3)}         &
\colhead{(4)}         &
\colhead{(5)}         &
\colhead{(6)}         &
\colhead{(7)}         &
\colhead{(8)}         &
\colhead{(9)}         &
\colhead{(10)}         &
\colhead{(11)}         &
\colhead{(12)}         &
\colhead{(13)}       &
\colhead{(14)}           \\
\noalign{\smallskip}\hline\noalign{\smallskip}
\multicolumn{14}{c}{Chandra Cycle 17 Objects}
}
  \startdata
$       080040.77+391700.4$&$  -27.93$&$  5.16$&$ 2.70_{-0.94}^{+1.34}$&$         1.84$&$    6.58$&$    44.77$&$   3.57$&$        31.47$&$      -1.82$&$    -0.12(0.78)$&$   2.01$&$    -0.09$&$  <1.3$\\
$       095023.19+024651.7$&$  -27.48$&$  3.66$&$                        <0.85$&$        <0.52$&$  <1.75$&$ <44.32$&$   1.82$&$        31.22$&$     <-1.92$&$ <-0.26(1.59)$&$ >4.74$&$     0.03$&$  <2.7$\\
$       101209.62+324421.4$&$  -27.77$&$  1.78$&$ 1.75_{-0.65}^{+0.95}$&$         1.06$&$    3.09$&$    44.73$&$   2.42$&$        31.38$&$      -1.88$&$    -0.19(1.23)$&$   3.12$&$     0.12$&$  <1.8$\\
$       101945.26+211911.0$&$   -28.08$&$  2.11$&$ 3.73_{-1.11}^{+1.50}$&$         2.46$&$    9.70$&$    44.89$&$   3.51$&$        31.49$&$      -1.75$&$    -0.05(0.31)$&$   1.32$&$   -0.05$&$   <1.3$\\
$       110409.96+434507.0$&$   -27.29$&$  1.21$&$                       <0.53$&$        <0.31$&$  <1.02$&$  <44.05$&$  1.75$&$         31.17$&$    <-2.01$&$ <-0.35(2.11)$&$  >8.15$&$    0.11$&$    <2.5$\\
$       122048.52+044047.6$&$   -27.73$&$ 1.65$&$2.16_{-0.80}^{+1.17}$&$          1.44$&$    5.71$&$    44.57$&$   2.69$&$        31.33$&$      -1.79$&$    -0.11(0.72)$&$   1.97$&$   -0.10$&$   <1.7$\\
$       122311.28+124153.9$&$  -27.81$&$   2.52$&$                       <0.89$&$        <0.54$&$  <1.85$&$  <44.36$&$  2.05$&$         31.34$&$    <-1.94$&$ <-0.25(1.62)$&$ >4.57$&$   0.14$&$   <3.2$\\
$       122855.90+341436.9$&$  -28.45$&$   1.33$&$                      <1.05$&$        <0.62$&$  <2.16$&$  <44.51$&$   3.29$&$         31.58$&$   <-1.99$&$ <-0.27(1.88)$&$  >5.15$&$   0.40$&$    <1.4$\\
$       124516.46+015641.1$&$  -27.70$&$  1.75$&$2.19_{-0.72}^{+1.01}$&$         1.31$&$    3.40$&$   44.87$&$    2.41$&$        31.39$&$      -1.86$&$   -0.17(1.11)$&$      2.80$&$   0.04$&$   <1.9$\\
$       140701.59+190417.9$&$  -27.67$&$  2.41$&$0.40_{-0.27}^{+0.55}$&$         0.26$&$    0.90$&$    44.08$&$    1.78$&$        31.26$&$     -2.03$&$    -0.36(2.24)$&$    8.74$&$   0.03$&$   <2.6$\\
$       150921.68+030452.7$&$  -28.40$&$  3.85$&$6.10_{-1.43}^{+1.81}$&$         3.94$&$  13.09$&$    45.16$&$    5.34$&$        31.65$&$      -1.77$&$   -0.04(0.30)$&$     1.29$&$   0.05$&$   <0.9$\\
$       155621.31+112433.2$&$  -28.16$&$  3.69$&$2.37_{-0.87}^{+1.28}$&$         1.47$&$    3.86$&$    44.81$&$    5.11$&$        31.60$&$      -1.97$&$   -0.25(1.70)$&$    4.39$&$   0.14$&$   <0.9$\\
$       163810.07+115103.9$&$  -27.73$&$   4.56$&$                      <0.59$&$        <0.37$&$   <1.25$&$ <44.21$&$    2.52$&$         31.40$&$  <-2.04$&$ <-0.34(2.24)$&$  >7.90 $&$   0.04$&$   <1.9$\\
$       232519.33+011147.8$&$  -27.15$&$  4.40$&$1.71_{-0.63}^{+0.93}$&$         1.11$&$    3.07$&$     44.66$&$   1.82$&$         31.15$&$     -1.83$&$   -0.18(1.05)$&$     2.87$&$   -0.03$&$  <2.4$\\
  \noalign{\smallskip}\hline\noalign{\smallskip}
  \multicolumn{14}{c}{Archival data} \\
  \noalign{\smallskip}\hline\noalign{\smallskip}
$       123326.03+451223.0$&$   -28.53$&$  1.36$&$47.9_{-13.9}^{+13.9}$\tablenotemark{a}&$ 28.81$&$ 122.43$&$   46.01$&$   4.80$&$        31.67$&$      -1.38$&$    0.35(2.44)$&$     0.12$&$    -0.10$&$      3.2$\\
$        215954.45-002150.1$&$   -29.20$&$  5.10$&$13.66_{-1.35}^{+1.49}$&$        6.51$&$  27.64$&$   45.36$&$    7.94$&$        31.89$&$      -1.71$&$     0.05(0.35)$&$    0.75$&$     0.07$&$     2.8$\\
$       113949.39+460012.9^*$&$  -27.38$&$  2.26$&$471.14_{-241.54}^{+241.54}$\tablenotemark{b}&$       48.07$&$ 197.62$&$   46.17$&$   1.92$&$       31.23$&$      -1.15$&$     0.52(3.21)$&$     0.04$&$   -0.30$&$    3.5$

  \enddata
  \tablecomments{
Col. (1): Object name. The object marked with ``$^*$'' comes from the extreme subsample.
Col. (2): Absolute $i$-band magnitude.
Col. (3): The column density of Galactic neutral hydrogen \citep{Dickey1990}.
Col. (4): Soft-band (0.5--2~keV) count rate in units of 10$^{-3}$~s$^{-1}$.
Col. (5): Observed-frame 0.5--2~keV flux (corrected for Galactic absorption) in units of $10^{-14}$~\flux.
Col. (6): Flux density at rest-frame 2~keV in units of $10^{-32}$~\mflux.
Col. (7): Logarithm of the rest-frame 2--10~keV luminosity in units of \lum, derived from $\Gamma_{\rm eff}$ and the observed-frame 0.5--2~keV flux.
Col. (8): Flux density at rest-frame 2500~\AA\ in units of $10^{-27}$~\mflux.
Col. (9): Logarithm of the rest-frame 2500~\AA\ luminosity density in units of \mlum.
Col. (10): Observed $\alpha_{\rm OX}$.
Col. (11): The difference between the observed $\alpha_{\rm OX}$ and the expectation from the
\hbox{$\alpha_{\rm OX}$--$L_{\rm 2500~{\textup{\AA}}}$} relation \citep{Just2007}. 
In the parentheses, the statistical significance of this difference is presented in units of the $\alpha_{\rm OX}$ rms scatter from Table~5 of Steffen et al. (2006).
Col. (12): The factor of X-ray weakness.
Col. (13): Relative SDSS $g-i$ color.
Col. (14): Radio-loudness parameter.\newline
$\rm ^a$For \textit{RASS2RXS} data, the count rate is for 0.1--2.4~keV.\newline
$\rm ^b$For \textit{XMMSL2} data, the count rate is for 0.2--2~keV.
}
  \label{aoxtable}
 \end{deluxetable}
\end{landscape}

\begin{deluxetable}{ccccccccccc}
\tabletypesize{\footnotesize}
\setlength{\tabcolsep}{4pt}
 \tablewidth{0pt}
 \tablecaption{Stacked X-ray Properties of X-ray Weak Quasars in the Representative Sample}
 \tablehead{
\colhead{Sample} &
\colhead{$N_{\rm weak}$} &
\colhead{Mean} &
\colhead{Total Stacked} &
\colhead{Soft-Band} &
\colhead{Hard-Band} &
\colhead{$\Gamma_{\rm eff}$}  &
\colhead{$\alpha_{\rm OX}$}                   &
\colhead{$\Delta\alpha_{\rm OX}(\sigma)$}    &                      
\colhead{$s$($\Delta\alpha_{\rm OX}$)} \\
\colhead{} &
\colhead{} &
\colhead{Redshift} &
\colhead{Exposure (ks)} &
\colhead{Counts} &
\colhead{Counts} &
\colhead{}   &                   
\colhead{}   &
\colhead{}  &
\colhead{} \\
\colhead{(1)}         &
\colhead{(2)}         &
\colhead{(3)}         &
\colhead{(4)}         &
\colhead{(5)}         &
\colhead{(6)}         &
\colhead{(7)}         &
\colhead{(8)}         & 
\colhead{(9)}         &
\colhead{(10)}
}
 \startdata
Extreme  &   7 & 1.823  & 22.9&         $  <3.9$&                            $<2.5$&                         $...$&$ <-2.33$&$ <-0.64(4.17)$ & ... \\
Bridge   &   7 &  1.945 & 30.0&         $ 12.2^{+4.8}_{-3.6}$&     $11.5^{+4.9}_{-3.6}$& $ 1.09^{+0.55}_{-0.45}$&$ -2.14 $&$ -0.45(2.93)$ &  0.11 \\ 
Representative & 14 &  1.884 & 52.9&$13.2^{+4.9}_{-3.7}$&     $11.2^{+4.9}_{-3.6}$& $ 1.19^{+0.56}_{-0.45}$&$  -2.22$&$ -0.53(3.44)$ &   0.10
 \enddata
 \tablecomments{
Col. (1): Sample or subsample name.
Col. (2): Number of X-ray weak WLQs in the sample (or subsample).
Col. (3): Mean redshift of X-ray weak WLQs in the sample (or subsample).
Col. (4): The total \chandra\ exposure time of X-ray weak WLQs in the sample (or subsample).
Cols. (5) and (6): Aperture-corrected stacked net counts in the soft (0.5--2~keV) and hard (2--8~keV) bands. 
If the source is not detected, an upper limit at a 90\% confidence level is listed.  
Col. (7): See the note of Col. (11) in Table \ref{xraytable}.
Cols. (8)--(9): See the note of Cols. (10)--(11) in Table \ref{aoxtable}.
Col. (10): Estimated error of the stacked \daox\ based on bootstrapping counts. 
}
 \label{stacktable}
\end{deluxetable}

\begin{deluxetable}{lccccccccc}
\tabletypesize{\footnotesize}
\setlength{\tabcolsep}{4pt}
 \tablewidth{0pt}
 \tablecaption{UV Emission-Line Measurements of Quasars in the Bridge Subsample from the SDSS Spectra}
 \tablehead{
\colhead{Object Name}                   &
\colhead{MJD}                  &
\colhead{\iona{C}{iv} Blueshift}                  &
\colhead{\iona{C}{iv} FWHM}                   &
\colhead{REW}                   &
\colhead{REW}                   &
\colhead{REW}                   &
\colhead{REW}                   &
\colhead{REW}                   &
\colhead{REW}                   \\
\colhead{(J2000)}   &
\colhead{}   &
\colhead{(km~s$^{-1}$)}   &
\colhead{(km~s$^{-1}$)}   &
\colhead{\iona{C}{iv}}   &
\colhead{\iona{Si}{iv}}   &
\colhead{$\lambda1900$~\AA}   &
\colhead{\iona{Fe}{ii}}   &
\colhead{\iona{Fe}{iii}}   &
\colhead{\iona{Mg}{ii}}   \\
\colhead{(1)}         &
\colhead{(2)}         &
\colhead{(3)}         &
\colhead{(4)}         &
\colhead{(5)}         &
\colhead{(6)}         &
\colhead{(7)}         &
\colhead{(8)}         &
\colhead{(9)}         &
\colhead{(10)}  \\
\noalign{\smallskip}\hline\noalign{\smallskip}
\multicolumn{9}{c}{Chandra Cycle 17 Objects}
}
  \startdata
$       080040.77+391700.4$&$    55528$&$  -2384\pm 114$&$   7058\pm 565$&$      13.5\pm 0.5$&$  8.6\pm 0.6$&$  15.8\pm 0.6$&$   24.5\pm 1.2$&$5.5\pm 0.6$&$ 21.9\pm 1.1$\\
$       095023.19+024651.7$&$    51908$&$  -3675\pm 834$&$  5380\pm 421$&$        8.0\pm 1.7$&$   7.5\pm 2.1$&$  17.1\pm 2.0$&$   33.5\pm 3.6$&$5.0\pm 1.8$&$ 23.9\pm 4.1$\\
$       101209.62+324421.4$&$    53442$&$ -2865\pm 419$&$   8497\pm 627$&$      12.5\pm 1.6$&$   8.0\pm 1.8$&$  18.3\pm 1.9$&$   31.7\pm 6.0$&$5.2\pm 1.4$&$ 23.2\pm 3.6$\\
$       101945.26+211911.0$&$    56027$&$     -155\pm 43$&$   2352\pm 204$&$      13.1\pm  0.6$&$   5.1\pm 0.7$&$   11.7\pm 0.8$&$   10.0\pm 1.5$&$4.1\pm 0.8$&$ 17.5\pm 1.2$\\
$       110409.96+434507.0$&$    53053$&$   -483\pm 279$&$   5187\pm 415$&$      10.8\pm  1.5$&$ 14.2\pm 2.2$&$  13.5\pm 1.5$&$   28.5\pm 2.7$&$4.7\pm 1.6$&$ 26.0\pm 2.9$\\
$       122048.52+044047.6$&$    52378$&$ -3488\pm 444$&$  10361\pm 793$&$       12.1\pm1.4$&$               ...$&$  15.5\pm 1.4$&$   28.5\pm 2.4$&$5.6\pm 1.7$&$ 28.7\pm 2.2$\\
$       122311.28+124153.9$&$    56308$&$ -3547\pm 531$&$    9531\pm 730$&$      7.8\pm   0.9$&$   5.8\pm 0.9$&$  11.3\pm 1.2$&$  29.1\pm  2.4$&$4.0\pm 1.1$&$  20.3\pm 2.7$\\
$       122855.90+341436.9 $&$    55571$&$   -972\pm 107$&$   2979\pm 239$&$        7.0\pm 0.6$&$   4.3\pm 0.6$&$    8.0\pm 0.9$&$  18.5\pm  1.8$&$1.7\pm 0.7$&$  11.5\pm 1.8$\\
$       124516.46+015641.1$&$    52024$&$ -3069\pm 245$&$   7876\pm 622$&$       13.3\pm 1.0$&$  7.4\pm 1.0$&$  18.4\pm 1.6$&$   21.5\pm 3.1$&$3.0\pm 1.1$&$ 25.8\pm 3.4$\\
$       140701.59+190417.9$&$    56039$&$ -5723\pm 397$&$ 10651\pm 798$&$       8.4\pm 1.0$&$  5.5\pm 0.9$&$  18.7\pm 1.6$&$   27.6\pm 2.9$&$2.6\pm 1.2$&$ 23.8\pm 2.5$\\
$       150921.68+030452.7$&$    55652$&$ -1869\pm 112$&$   6715\pm 517$&$       15.5\pm 0.7$&$  6.2\pm 0.8$&$  18.2\pm 0.8$&$   23.3\pm 1.5$&$4.8\pm 0.8$&$  20.9\pm 1.2$\\
$       155621.31+112433.2$&$    54572$&$      233\pm 87$&$   2834\pm 227$&$       9.3\pm 0.8$&$                ...$&$   10.7\pm 0.8$&$  19.6\pm 1.5$&$1.6\pm 0.8$&$ 20.2\pm 1.4$\\
$       163810.07+115103.9$&$    54585$&$ -1153\pm 141$&$   4485\pm 360$&$       11.1\pm 0.9$&$  7.9\pm 1.0$&$   10.7\pm 1.2$&$  13.8\pm 2.1$&$1.4\pm 0.9$&$  18.7\pm 2.4$\\
$        232519.33+011147.8$&$   51818$&$ -5208\pm 523$&$   6564\pm 526$&$       7.4\pm 1.9$&$                ...$&$   12.8\pm 1.7$&$  30.0\pm 3.1$&$5.8\pm 2.0$&$  25.4\pm 3.1$\\
 \noalign{\smallskip}\hline\noalign{\smallskip}
 \multicolumn{9}{c}{Archival Data} \\
 \noalign{\smallskip}\hline\noalign{\smallskip}
$       123326.03+451223.0$&$    56367$&$  -1091\pm 61$&$   3877\pm 310$&$      10.9\pm 0.6$&$   5.8\pm 0.5$&$  14.4\pm 0.8$&$   21.8\pm 1.6$&$3.6\pm 0.7$&$ 19.8\pm 1.3$\\
$        215954.45-002150.1$&$    52173$&$    -163\pm 65$&$   2698\pm 216$&$     7.4\pm 0.6$&$    4.6\pm 0.6$&$    8.9\pm 0.8$&$   15.8\pm 1.4$&$1.5\pm 0.6$&$ 13.8\pm 1.3$\\
$       113949.39+460012.9^*$&$    56398$&$-3408\pm 413$&$   7808\pm 656$&$      5.7\pm 0.7$&$      5.0\pm 0.7$&$    9.4\pm 1.2$&$  24.9\pm 2.0$&$2.1\pm 1.0$&$13.5\pm 1.8$
  \enddata
 \tablecomments{
Col. (1): Object name. The object marked with ``$^*$'' comes from the extreme subsample.
Col. (2): Modified Julian date of the SDSS observation.
Cols. (3)--(5): Blueshift, full width at half maximum (FWHM), and REW (in units of \AA) of \iona{C}{iv}~$\lambda1549$.
Cols. (6)--(10): REWs (in units of \AA) of the \iona{Si}{iv}~$\lambda1397$, $\lambda1900$ complex, \iona{Fe}{ii} (2250--2650~\AA), \iona{Fe}{iii} UV48 $\lambda2080$, and \iona{Mg}{ii}~$\lambda2799$ emission features. 
The \iona{Si}{iv}~$\lambda1397$ feature is a mixture of 
\iona{Si}{iv} and \iona{O}{iv]}.
The $\lambda1900$ complex mainly consists of \iona{C}{iii]}~$\lambda1909$, and it also includes
\iona{[Ne}{iii]}~$\lambda1814$, \iona{Si}{ii}~$\lambda1816$, 
\iona{Al}{iii}~$\lambda1857$, and \iona{Si}{iii]}~$\lambda1892$.
``...'' indicates insufficient spectral coverage for accurate measurements.
}
 \label{uvtable}
\end{deluxetable}

\clearpage
\begin{deluxetable}{@{\extracolsep{4pt}}ccccccccc}
\tabletypesize{\footnotesize}
\setlength{\tabcolsep}{2pt}
\tablewidth{0pt}
 \tablecaption{X-ray Weakness Properties of Quasar Subsamples}
 \tablehead{\colhead{} & \multicolumn{3}{c}{Sample Properties} & \multicolumn{4}{c}{$P_{\rm null}$ of Fisher's Exact Test} \\
 \cline{2-4}  \cline{5-9}
 \colhead{Sample} & \colhead{Median} & \colhead{$N_{\rm weak}$} & \colhead{$F_{\rm weak}$} & \colhead{Extreme} & \colhead{Bridge}& \colhead{\citet{Gibson2008}}&\colhead{\citet{Gibson2008}} \\
 \colhead{Name} & \colhead{$\Delta\alpha_{\rm OX}$}& \colhead{} & \colhead{} & \colhead{Subsample} & \colhead{Subsample} & \colhead{Subsample 1} & \colhead{Subsample 2}\\
 \colhead{} & \colhead{(1)} & \colhead{(2)} & \colhead{(3)} & \colhead{(4)} & \colhead{(5)}& \colhead{(6)} & \colhead{(7)}}
 
 \startdata
 Extreme &   $-0.10^{+0.12}_{-*}$ &  7 & $       43.8\%^{+12.4\%}_{-11.0\%}$&  1  & 1  & $5\times 10^{-4}$& $1\times 10^{-4}$&\\
 \vspace{1mm}Subsample &&&&&&&&\\
 Bridge &   $-0.19^{+0.04}_{-*}$ &   7 & $      43.8\%^{+12.4\%}_{-11.0\%}$&   1 &  1      & $5\times 10^{-4}$&  $1\times 10^{-4}$ &\\ 
 \vspace{1mm}Subsample&&&&&&&&\\
 \citet{Gibson2008} &  $-0.02^{+0.01}_{-0.02}$  &  3 & $     5.1\%^{+4.5\%}_{-1.6\%}$&          $5\times 10^{-4}$  &  $5\times 10^{-4}$      &    1   &   0.68&\\ 
 \vspace{1mm}Subsample 1&&&&&&&&\\
 \citet{Gibson2008}  & $0.05^{+0.02}_{-0.02}$  & 2 & $ 3.3\%^{+4.1\%}_{-1.1\%}$&                 $1\times 10^{-4}$ & $1\times 10^{-4}$  &   0.68 & 1&\\
 Subsample 2 &&&&&&&&
 \enddata
 \tablecomments{
Col. (1): Median $\Delta\alpha_{\rm OX}$ values of each subsample. The $1\sigma$ uncertainties of $\Delta\alpha_{\rm OX}$ are estimated via bootstrap. *For the extreme subsample and the bridge subsample, the $1\sigma$ lower bound of median $\Delta\alpha_{\rm OX}$ cannot be appropriately estimated as a result of having too many upper limits in the sample with unknown distributions.
Col. (2): Number of X-ray weak quasars in each subsample.
Col. (3): The fraction of X-ray weak quasars with uncertainties calculated following \citet{Cameron2011}.
Cols. (4)--(7): Probabilities of the two corresponding subsamples having an equal fraction of X-ray weak quasars using Fisher's exact test.
}
 \label{fractable}
\end{deluxetable}

\begin{deluxetable}{@{\extracolsep{8pt}}ccccccccc}
\tabletypesize{\footnotesize}
\setlength{\tabcolsep}{12pt}
\tablewidth{0pt}
 \tablecaption{Results from Peto-Prentice Tests Comparing the Spectral Properties of X-ray Weak and X-ray Normal WLQs}
 \tablehead{
 \colhead{Spectral} &
 \multicolumn{4}{c}{Representative Sample} &
 \multicolumn{4}{c}{Full sample} \\
 \cline{2-5}  \cline{6-9}
 \colhead{Property} &\colhead{$N_{\rm weak}$} &\colhead{$N_{\rm normal}$} &\colhead{$\sigma$} &\colhead{$P_{\rm null}$} &\colhead{$N_{\rm weak}$} &\colhead{$N_{\rm normal}$} &\colhead{$\sigma$} & \colhead{$P_{\rm null}$} 
 }
 \startdata 
 \iona{C}{iv} REW                 & 14  & 18  &     0.3    &$0.79$    &  26  &  20 &   0.03  &$0.98$\\ 
\iona{C}{iv} blueshift            & 14  & 18  &     0.4    &$0.67$    & 26   & 19  &  1.4    &$0.16$\\
\iona{C}{iv} FWHM              &  14  & 18  &   0.08   &$0.94$    & 26  &  19  &  1.0   &$0.31$\\ 
\iona{Si}{iv} REW                &   9   &  13  &     0.3   &$0.74$    & 24  &  19  &   0.06   &$0.95$\\
$\lambda1900$ \AA\ REW  &  14  &  18  &     0.6    &$0.57$    & 32  &  26  &   0.2  &$0.86$\\
\iona{Fe}{ii} REW                &  14  &  18  &     1.5    &$0.14$    & 32  &  28  &   3.1  &$2\times 10^{-3}$\\
\iona{Fe}{iii} REW                & 14  &  18  &      0.9    &$0.38$    & 32  &  28  &  0.9  &$0.35$\\
\iona{Mg}{ii} REW               &  14  &  16  &      0.5    &$0.65$    & 27  & 25  &   0.7   &$0.47$\\  
$\Delta(g-i)$                        &  14  &  18  &     2.4     &$0.01$    & 33  & 26  &   3.9  &$1\times 10^{-4}$ 
 \enddata
 \tablecomments{The Peto-Prentice test was performed for both the representative sample and the full sample.
It reduces to Gehan's Wilcoxon test when there is no censoring. 
For each spectral property, we list the number of X-ray weak WLQs ($N_{\rm weak}$) and the number of X-ray normal WLQs ($N_{\rm normal}$) available for the test, the test statistic ($\sigma$), and the probability of the distributions of this spectral property among X-ray weak WLQs and X-ray normal WLQs being drawn from the same population.}
 \label{pptable}
\end{deluxetable}


\bsp	
\label{lastpage}
\end{document}